\documentclass{article}

\usepackage{arxiv}

\usepackage{pgfplots}
\usepackage{subfigure}
\usepackage[utf8]{inputenc} % allow utf-8 input
\usepackage[T1]{fontenc}    % use 8-bit T1 fonts
\usepackage{hyperref}       % hyperlinks
\usepackage{url}            % simple URL typesetting
\usepackage{booktabs}       % professional-quality tables
\usepackage{amsfonts}       % blackboard math symbols
\usepackage{nicefrac}       % compact symbols for 1/2, etc.
\usepackage{microtype}      % microtypography
\usepackage{lipsum}
\usepackage{cite}
\usepackage{amsmath}
\usepackage{amssymb}

\usepackage{bm}  % jps: for bold-math 

\title{Deep learning approaches to surrogates for solving the diffusion equation for mechanistic real-world simulations}

\author{
  J. Quetzalc\'oatl Toledo-Mar\'in \\
  Biocomplexity Institute, Indiana University, \\
  Bloomington, IN 47408, USA\\
  \texttt{j.toledo.mx@gmail.com} \\
  \And
  Geoffrey Fox \\
  Digital Science Center, \\
  Bloomington, IN 47408, USA \\
  Luddy School of Informatics, \\
  Computing and Engineering, IN 47408, USA \\
  \texttt{gcf@iu.edu} \\
  \And
 James Sluka \\
  Biocomplexity Institute, Indiana University, \\
  Bloomington, IN 47408, USA\\
  \texttt{jsluka@iu.edu} \\
  \And
 James A. Glazier \\
  Biocomplexity Institute, Indiana University, \\
  Bloomington, IN 47408, USA\\
  \texttt{jaglazier@gmail.com} \\
}

\begin{document}
\maketitle

% \textcolor{red}{jps: Changes of note highlighted in red. Please double check. Search for "textcolor".}

% \textcolor{red}{jps: I think we should replace "NN 1" and "NN 2" with NN1 and NN2.}

\begin{abstract}
    	In many mechanistic medical, biological, physical and engineered spatiotemporal dynamic models the numerical solution of partial differential equations (\textbf{PDEs}), especially for diffusion, fluid flow and mechanical relaxation, can make simulations impractically slow. Biological models of tissues and organs often require the simultaneous calculation of the spatial variation of concentration of dozens of diffusing chemical species. One clinical example where rapid calculation of a diffusing field is of use is the estimation of oxygen gradients in the retina, based on imaging of the retinal vasculature, to guide surgical interventions in diabetic retinopathy. Since the quasi-steady-state solutions required for fast-diffusing chemical species like oxygen are particularly computationally costly, we consider the use of a neural network to provide an approximate solution to the steady-state diffusion equation. Machine learning \textbf{surrogates}, neural networks trained to provide approximate solutions to such complicated numerical problems, can often provide speed-ups of several orders of magnitude compared to direct calculation. Surrogates of PDEs could enable use of larger and more detailed models than are possible with direct calculation and can make including such simulations in real-time or near-real time workflows practical. Creating a surrogate requires running the direct calculation tens of thousands of times to generate training data and then training the neural network, both of which are computationally expensive. Often the practical applications of such models require thousands to millions of replica simulations, for example for parameter identification and uncertainty quantification, each of which gains speed from surrogate use and rapidly recovers the up-front costs of surrogate generation.  We use a Convolutional Neural Network to approximate the stationary solution to the diffusion equation in the case of two equal-diameter, circular, constant-value sources located at random positions in a two-dimensional square domain with absorbing boundary conditions. Such a configuration caricatures the chemical concentration field of a fast-diffusing species like oxygen in a tissue with two parallel blood vessels in a cross section perpendicular to the two blood vessels. To improve convergence during training, we apply a training approach that uses roll-back to reject stochastic changes to the network that increase the loss function. The trained neural network approximation is about 1000 times faster than the direct calculation for individual replicas. Because different applications will have different criteria for acceptable approximation accuracy, we discuss a variety of loss functions and accuracy estimators that can help select the best network for a particular application. We briefly discuss some of the issues we encountered with overfitting, mismapping of the field values and the geometrical conditions that lead to large absolute and relative errors in the approximate solution.  
\end{abstract}

% keywords can be removed

\section{Introduction}
Diffusion is ubiquitous in physical, biological and engineered systems. In mechanistic computer simulations of the dynamics of such systems, solving the steady state and time-varying diffusion equations with multiple sources and sinks is often the most computationally expensive part of the calculation, especially in cases with multiple diffusing species with diffusion constants differing by multiple orders of magnitude. Examples in biology include cells secreting and responding to diffusible chemical signals during embryonic development, blood vessels secreting oxygen which cells in tissues absorb during normal tissue function, tumors secreting growth factors promoting neoangiogenesis in cancer progression, or viruses spreading from their host cells to infect other cells in tissues. In these situations the natural diffusion constants can range from $\sim 10^{3} \text{$\mu$m}^2/\text{s}$ for oxygen to $\sim 0.1-10^{2} \text{$\mu$m}^2/\text{s}$ for a typical protein \cite{phillips2018membranes}. Dynamic simulations of biological tissues and organs may require the independent calculation of the time-varying concentrations of dozens of chemical species in three dimensions, and in the presence of a complex field of cells and extracellular matrix. As the number of species increases, solving these diffusion equations dominates the computational cost of the simulation. Numerous approaches attempt to reduce the cost of solving the diffusion equation including implicit, particle-based, frequency-domain and finite-element methods, multithreaded and MPI-based parallelization and GPUs, but all have significant limitations. 
%as well as data-driven methods for discovering the underlying governing equation of a given process as in the case of SINDy \cite{champion2019data}. 
In real-world problems, the number of sources and sinks, their shape,  boundary fluxes and positions differ from instance to instance and may change in time. Boundary conditions may also be complicated and diffusion constants may be anisotropic or vary in space. The resulting lack of symmetry means that many high-speed implicit and frequency-domain diffusion-solver approaches do not work effectively, requiring the use of simpler but slower forward solvers \cite{schiesser2012numerical}. Deep learning \footnote{We use the terms \textit{deep learning} and \textit{machine learning} interchangeably. We also use \textit{neural network} and \textit{deep neural network} interchangeably.} surrogates to solve either the steady-state field or the time-dependent field for a given set of sources and sinks subject to diffusion could potentially increase the speed of such simulations by several orders of magnitude compared to the use of direct numerical solvers.  

\begin{figure}[hbtp]
\centering
\centering
\subfigure[Initial condition]{
\includegraphics[width=3.05in, height=3.1in]{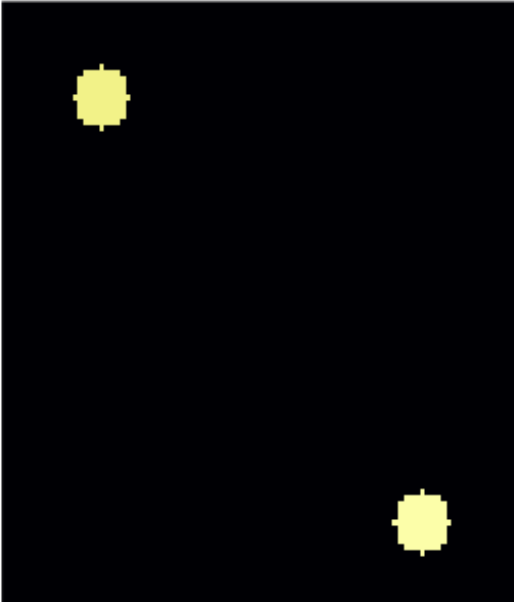}
\label{fig:in}}
\centering
\subfigure[Stationary solution]{
\includegraphics[width=3.1in, height=3.1in]{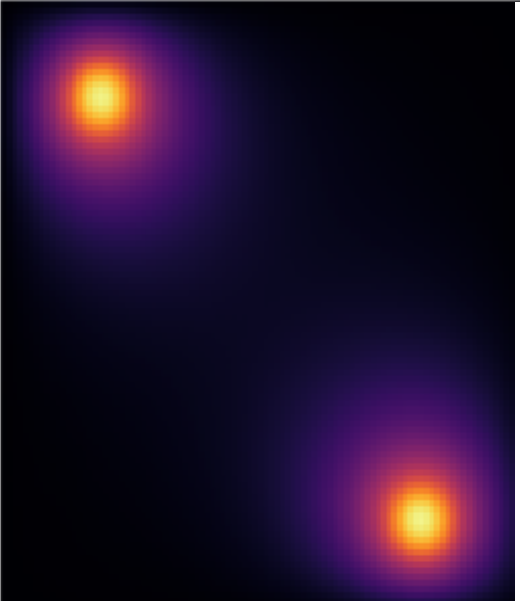}
\label{fig:out}}
\caption{ Snapshot of \textbf{a)} initial condition and \textbf{b)} stationary state solution. \textbf{a)} We placed two random value sources of radius $5\;voxels$ in random positions fully within a $100\times100 pixel$ lattice and used this configuration as the input to the NN. \textbf{b)} Stationary solution to the diffusion equation with absorbing boundary conditions for the initial conditions in \textbf{a)}. The stationary solution \textbf{b)} is the target for the NN. We fixed the diffusion constant to $D = 1\;voxels^2/s$ and the decay rate to $\gamma = 1/400s^{-1}$, which yields a diffusion length equal to $\sqrt{D/\gamma}\;voxels = 20 voxels$.
} \label{fig:inout0}
\end{figure}

One challenge in developing effective deep neural network (NN) diffusion-solver surrogates is that the dimensionality of the problem specification is potentially very high, with an arbitrary pattern of sources and sinks, with different boundary conditions for each source and sink, and spatially variable or anisotropic diffusivities. As a proof-of-principle we will start with a NN surrogate for a simple version of the problem that we can gradually generalize to a full surrogate in future work. 
%We first restrict to a two dimensional square domain with absorbing boundary conditions. We calculate the steady-state concentration field for a single round source {\bf describe the choice of fluxes and decay and reason for this choice} at an arbitrary spatial location. This would mimic, {\bf e.g.}, the concentration of oxygen in a tissue around a long capillary. In this case the key challenge is the interaction of the source with the boundary. 
In a two-dimensional square domain represented as $N \times N pixels$ and with absorbing boundary conditions, we place two circular sources of equal diameters at random positions, with the constraint that the sources do not overlap and are fully contained within the domain. Each source imposes a constant value on the diffusing field within the source and at its boundary. We select the value for one of the sources equal to 1 while the value for the other source is randomly selected from a uniform distribution between $(0,1]$ (see Fig. \ref{fig:in}). Outside the sources the field diffuses with a constant diffusion constant ($D$) and linearly decays with a constant decay rate ($\gamma$). This simple geometry could represent the diffusion and uptake of oxygen in a volume of tissue between two parallel blood vessels of different diameters. Although reflecting or periodic boundary conditions might better represent a potion of a larger tissue, we use the simpler absorbing boundary conditions here. In this case, the steady-state field depends critically on the distance between the sources, and between the sources and the boundary, both relative to the diffusion length ($l_{D} = (D/\gamma)^{1/2}$) and on the sources' field strengths. 

In practice then, the solution of the steady state diffusion equation maps an image consisting of $N \times N$ pixels with 0 value outside the sources and constant values between 0 and 1 inside the sources to a second image of the same size, which has the same values inside the sources but values between 0 and 1 elsewhere (see Fig. \ref{fig:out}).  We evaluate the ability of a NN trained on the explicit numerical solutions of the steady-state diffusion field for $20,000$ two-source examples to approximate the steady state field for configurations of sources that it had not previously encountered. 

Notice that the diffusion kernel convolution used in the direct solution of the time-dependent diffusion equation (e.g., finite-element methods) is a type of  convolutional neural network \cite{schiesser2012numerical}. Therefore we chose deep convolutional NN as the architecture. However, there are multiple types of convolutional NN. Here we considered two of these. A deep convolutional neural network and an autoencoder \cite{baur2020autoencoders}. In addition, because it was possible that these two types would  do better at replicating specific aspects of the overall solution, we also evaluated a superposition of the two. 
%We considered a number of neural network architectures for use in developing our NN surrogate diffusion solver.  
Time series surrogates often use recurrent NN \cite{zhang2000predicting, dubois2020data}. Similarly, deep generative models have been shown to be useful to sample from high dimensional space, as in the case of molecular dynamics and chemical reaction modeling \cite{chen2018molecular, zhang2019machine, noe2019boltzmann, gkeka2020machine, noe2020machine, kasim2020up}. Since our main interest is the stationary solution, we did not consider these approaches. 
%Since the classic forward time-dependent diffusion solver method applies a convolution kernel repeatedly to the field \cite{schiesser2012numerical}, convolutional neural networks could potentially mimic direct solvers. For instance, autoencoders are effective at generative models applied to images \cite{baur2020autoencoders}. 
%[Here we test whether autoencoders excel at this application. While technological developments have substantially decreased this by, for instance, parallelizing, the use of machine learning as surrogates for numerical solutions of partial differential equations is currently a highly active area] 
% \cite{chen2018molecular, zhang2019machine, noe2019boltzmann, gkeka2020machine, noe2020machine, kasim2020up, li2020fourier}.

%The sections are organized as follows: In the next section we present the neural network (NN) architecture and features. In section 3 we present our results and we discuss them. Section 4 is devoted for conclusions.

%[Why did you try the architectures you did?]

\section{Model}
\begin{figure}[hbtp]
% \centering
\centering
\subfigure[NN 1]{
\includegraphics[width=4.6in]{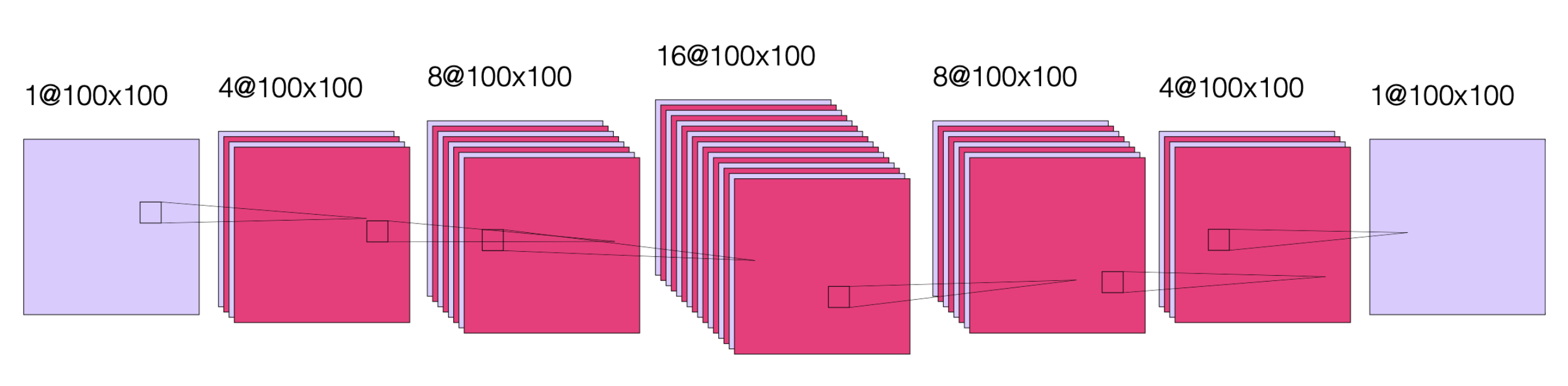}
\label{fig:NN1}}
\centering
\subfigure[NN 2]{
\includegraphics[width=4.6in]{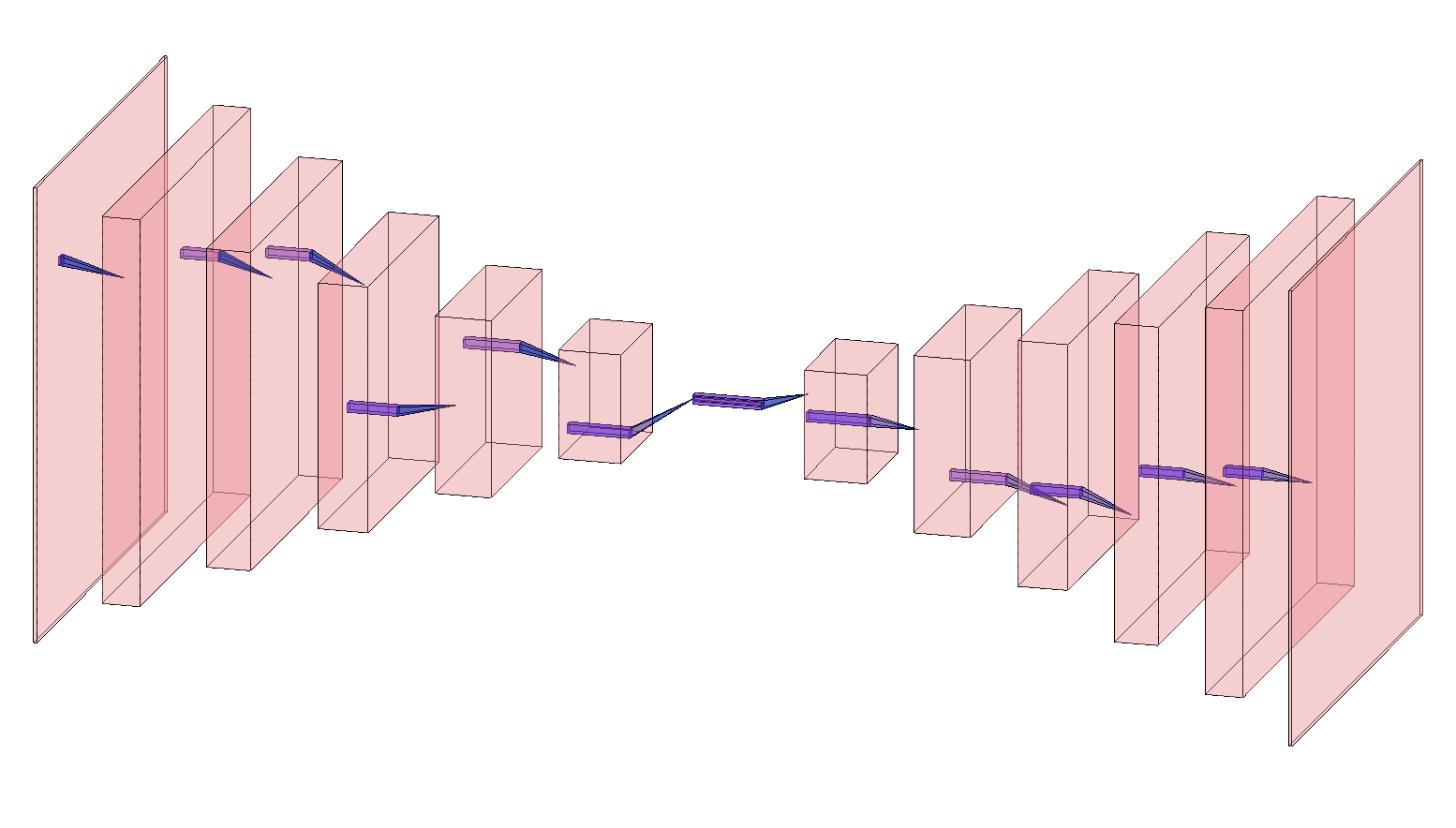}
\label{fig:NN2}}
\caption{ Sketch of \textbf{a)} Convolutional NN 1. The first layer takes as input a single-channel $N \times N$ image and applies four $3 \times 3 \time 1$ convolutions to generate four $N \times N$ images, the second layer applies eight $3 \times 3 \time 4$ convolutions to generate eight $N \times N$ images, the third layer applies sixteen $3 \times 3 \time 8$ convolutions to generate sixteen $N \times N$ images, the fourth layer applies eight $3 \times 3 \time 16$ convolutions to generate eight $N \times N$ images, the fifth layer applies four $3 \times 3 \time 8$ convolutions to generate four $N \times N$ images and the sixth layer applies a $3 \times 3 \time 4$ convolution to generate a single $N \times N$ image. Sketch of  \textbf{b)} autoencoder NN 2. The first 6 layers perform a meanpool operation that reduces image height and width by half after each layer, with the image dimensions following the sequence $\lbrace 100^2, 50^2, 25^2, 12^2, 6^2, 3^2, 1^2 \rbrace$ \textbf{while adding channels after each layer }following the sequence $\lbrace 1,64,128,256,512,1024,2048 \rbrace$. Then, the following 6 layers reverse the process, reducing the number of channels following the sequence $\lbrace 1024, 512, 256, 128, 64, 1 \rbrace$ while increasing the height and width following the sequence $\lbrace 1^2, 3^2, 7^2, 13^2, 25^2, 51^2, 100^2 \rbrace$.
%, whereas NN 2 is an autoencoder. 
This sketch only defines the kinds of layers used. For details about the activation functions used in each layer, see Table \ref{table:1}.
} \label{fig:NN}
\end{figure}

Fig. \ref{fig:NN} shows our NN architecture. We denote by $| x\rangle$ and $| \hat{y} \rangle$ the input and output images, that is the initial condition layout of the source cells and the predicted stationary solution of the diffusion equation, respectively. The input $| x\rangle$ passes to two different neural networks (NNs) denoted \textit{NN 1} (Fig. \ref{fig:NN1}) and \textit{NN 2} (Fig. \ref{fig:NN2}) which output $| \hat{y}_1 \rangle$ and $| \hat{y}_2 \rangle$, respectively. The output $| \hat{y} \rangle$ is a weighted sum of the outputs of the two NNs,  $|\hat{y} \rangle = p_1 | \hat{y}_1 \rangle + p_2 | \hat{y}_2 \rangle$, where $p_1$ and $p_2$ are fixed hyperparameters, \textit{i.e.}, these hyperparameters are fixed during training. In our code \cite{githubJaque} $p_i$ are real numbers, however, in this paper we only consider the Boolean case where they each take values of 0 or 1. \textit{NN 1} is a deep convolutional neural network that maintains the height and width of the input image through each of 6 convolutional layers. The first layer outputs a 4-channel image, the second layer outputs an 8-channel image, the third layer outputs a 16-channel image, the fourth layer outputs an 8-channel image, the fifth layer outputs a 4-channel image and the sixth layer outputs a 1-channel image. \textit{NN 2} is an autoencoder \cite{chen2017deep} where the first 6 layers perform a meanpool operation that reduces height and width in half after each layer following the sequence $\lbrace 100^2, 50^2, 25^2, 12^2, 6^2, 3^2, 1^2 \rbrace$ while adding channels after each layer following the sequence $\lbrace 1,64,128,256,512,1024,2048 \rbrace$. Then, the following 6 layers consist on reducing the number of channels following the sequence $\lbrace 1024, 512, 256, 128, 64, 1 \rbrace$ while increasing the height and width following the sequence $\lbrace 1^2, 3^2, 7^2, 13^2, 25^2, 51^2, 100^2 \rbrace$.  Fig. \ref{fig:NN} sketches the architectures of the two NNs, while Table \ref{table:1} provides their parameters. 
%The choice of this architecture was based on the type of problem setting, i.e., 
We will find that NN 1 will capture the sources whereas NN 2 will capture the field.
%, as we show in the next section. 
In Table \ref{table:1} we specify each neural network by specifying for each layer the kind of layer, the activation function and the output shape.
%the stationary solution of a diffusion-type equation which will contain sources that will be mainly captured by NN 1 and the field that will be captured by NN 2.

\begin{figure}[hbtp]
\centering
%\centering
%\subfigure[NN 1]{
\includegraphics[width=4.6in]{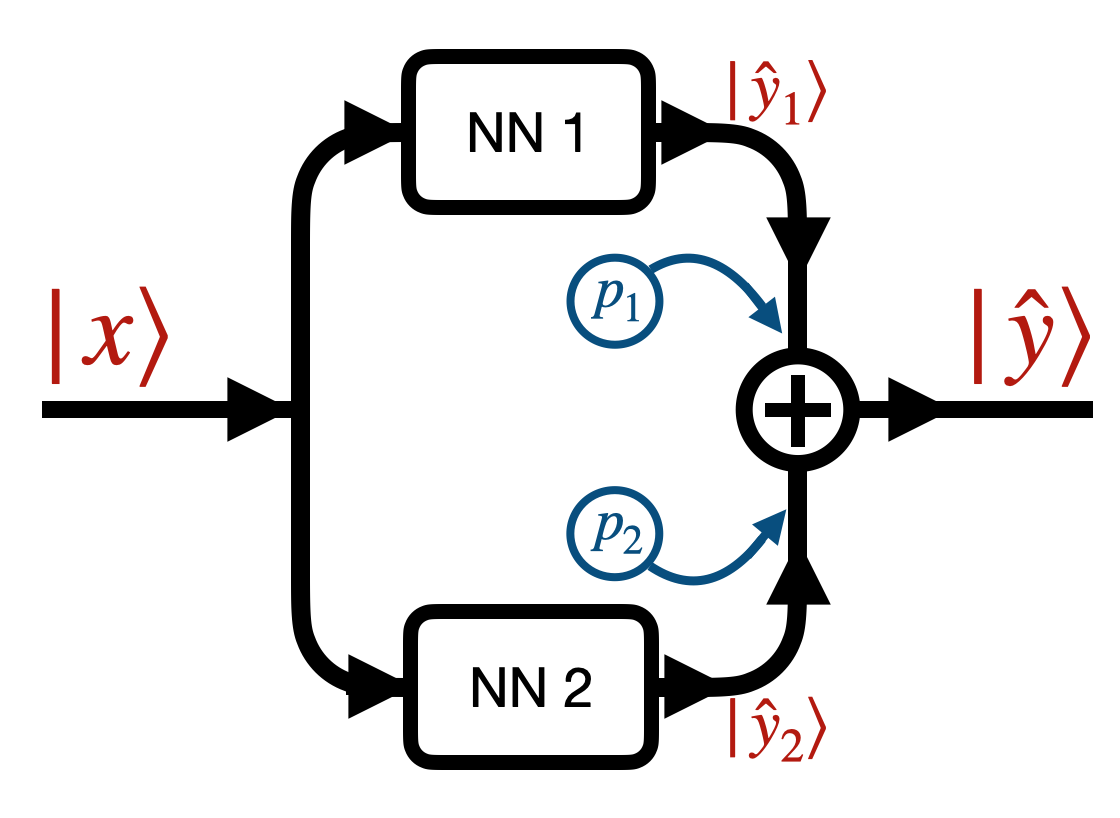}
\caption{ %Convolutional neural network. 
Network Architecture: The input image $|x\rangle$ passes through NN 1 ( see Fig. \ref{fig:NN1}) and NN 2 (see \ref{fig:NN2}), generating the two outputs $\hat{y}_1 \rangle $ and $ | \hat{y}_2 \rangle$ The final output $|\hat{y} \rangle$ is the sum of the  outputs of the two NNs weighted by coefficients $p_1$ and $p_2$, \textit{i.e.}, $|\hat{y} \rangle = p_1 | \hat{y}_1 \rangle + p_2 | \hat{y}_2 \rangle$.  $p_i$ are fixed Boolean hyperparameters for the model and fixed for each model we trained. This means that when a given model has $p_i=0$ ($p_i=1$) then $NN i$ is turned off (on).  } \label{fig:NN0}
\end{figure}

\begin{table}[h!]
\centering
{\renewcommand{\arraystretch}{1.3} % jps  add a bit of extra space above rows

\subfigure[NN 1]{
 \begin{tabular}{|c c c|} 
 \hline
 Operation & Act & Output shape \\ [0.5ex] 
 \hline\hline
 %Input & - & 1 x 100 x 100 \\ 
 %\hline
 Conv 3 x 3 & LReLU & 4 x 100 x 100 \\ 
 \hline
 Dropout 1 ($D_1$) & - & - \\
 \hline
 BatchNorm & Identity & - \\
 \hline
 Conv 3 x 3 & LReLU & 8 x 100 x 100 \\ 
 \hline
 BatchNorm & Identity & - \\
 \hline
 Conv 3 x 3 & LReLU & 16 x 100 x 100 \\ 
 \hline
 BatchNorm & Identity & - \\
 \hline
 Conv 3 x 3 & LReLU & 8 x 100 x 100 \\ 
 \hline
 BatchNorm & Identity & - \\
 \hline
 Conv 3 x 3 & LReLU & 4 x 100 x 100 \\ 
 \hline
 BatchNorm & Identity & - \\
 \hline
 Conv 3 x 3 & ReLU & 1 x 100 x 100 \\ 
 \hline
 Dropout 2 ($D_2$) & - & - \\
 \hline
 BatchNorm & Identity & - \\
 \hline
\end{tabular}}
\subfigure[NN 2]{
\begin{tabular}{|c c c|} 
 \hline
 Operation & Act & Output shape \\ [0.5ex] 
 \hline\hline
 %Input & - & 1 x 100 x 100 \\ 
 %\hline
 Conv 3 x 3 & LReLU & 64 x 100 x 100 \\ 
 \hline
 BatchNorm & Identity & - \\
 \hline
 Dropout 3 ($D_3$) & - & - \\
 \hline
 Meanpool & Identity & 64 x 50 x 50 \\
 \hline
 Conv 3 x 3 & LReLU & 128 x 50 x 50 \\ 
 \hline
 Meanpool & Identity & 128 x 25 x 25 \\
 \hline
 Conv 3 x 3 & LReLU & 256 x 25 x 25 \\ 
 \hline
 Meanpool & Identity & 256 x 12 x 12 \\
 \hline
 Conv 3 x 3 & LReLU & 512 x 12 x 12 \\ 
 \hline
 Meanpool & Identity & 512 x 6 x 6 \\
 \hline
 Conv 3 x 3 & LReLU & 1024 x 6 x 6 \\ 
 \hline
 Meanpool & Identity & 1024 x 3 x 3 \\
 \hline
 Conv 3 x 3 & LReLU & 2048 x 1 x 1 \\ 
 \hline
 ConvT 3 x 3 & LReLU & 1024 x 3 x 3 \\ 
 \hline
 ConvT 3 x 3 & LReLU & 512 x 7 x 7 \\ 
 \hline
 ConvT 3 x 3 & LReLU & 256 x 13 x 13 \\ 
 \hline
 ConvT 3 x 3 & LReLU & 128 x 25 x 25 \\ 
 \hline
 ConvT 3 x 3 & LReLU & 64 x 51 x 51 \\ 
 \hline
 Dropout 4 ($D_4$) & - & - \\
 \hline
 ConvT 4 x 4 & ReLU & 1 x 100 x 100 \\ 
 \hline
 BatchNorm & Identity & - \\
 \hline
\end{tabular}}

\caption{Convolutional Neural Network architectures. Left panel corresponds to the succesive operations of NN 1 while the right panel corresponds to the succesive operations NN 2. \textit{Act} stands for activation function. Conv, ConvT and (L)ReLU stand for convolution, convolution transpose, and (leaky) rectified linear unit, while Identity means the activation function is the identity function (see Ref. \cite{fluxWeb}). Both NNs take as input the initial condition which has dimensions $Channels \times Width \times Height = 1 \times 100 \times 100$ }
\label{table:1}
} % jps  add a bit of extra space above rows
\end{table}

To generate representative two-source initial conditions and paired steady-state diffusion fields, we considered a two-dimensional lattice of size $100\times100units^2$. We generated 20k configurations with two sources, each with a radius of $5\;units$. One source has a constant source value equal to $1$, while the other source has a constant source value between 0 and 1 randomly assigned using a uniform distribution.
%with \textcolor{red}{a source value for each randomly assigned uniform constant values between 0 and 1,} 
Everywhere else the field value is 0.  We placed the sources in randomly uniform positions in the lattice. This image served as the input for the NN $| x\rangle$. Then we calculated the stationary solution to the diffusion equation with absorbing boundary conditions for each initial condition using the \textit{Differential Equation} package in Julia \cite{rackauckas2017differentialequations}.
%When sources are near the boundary, absorbing boundary conditions create steep gradients, which can be challenging to the NN. Notice that analytically one can mimic absorbing boundary conditions by placing inverted images of the sources outside the lattice to guarantee that the field in the boundary is zero, whereas conventional NN's don't take this into consideration. 
The Julia-calculated stationary solution is the target or ground truth image for the NN $ | y \rangle$. In Figs. \ref{fig:in} and \ref{fig:out} we show an initial condition and the stationary solution, respectively.  
%In Figs. \ref{fig:input}, \ref{fig:target} and \ref{fig:output} we show 20 different inputs, targets and predictions, respectively. 
%These predictions were obtained using model 12. 
We have set the diffusion constant to $D = 1units^2/s$ and the decay rate $\gamma = 1/400s^{-1}$, which yield a diffusion length $l_D = \sqrt{D/\gamma}=20\; units$. Notice that this length is 4 times the radius of the sources and 1/5 the lattice linear dimension. As $\gamma$ increases and as $D$ decreases, this length decreases. As this length decreases, the field gradient also decreases \cite{tikhonov2013equations}. The source code to generate the data and train the NN can be found in Ref. \cite{githubJaque}.

% \begin{figure}[hbtp]
% \centering
% \centering
% \subfigure[Initial condition]{
% \includegraphics[width=3.05in, height=3.1in]{Figs/in.png}
% \label{fig:in}}
% \centering
% \subfigure[Stationary solution]{
% \includegraphics[width=3.1in, height=3.1in]{Figs/out.png}
% \label{fig:out}}
% \caption{ Snapshot of \textbf{a)} initial condition and \textbf{b)} stationary state solution. \textbf{a)} We placed two random value sources of radius $5\;voxels$ in random positions fully within a $100\times100 pixel$ lattice and used this configuration as the input to the NN. \textbf{b)} Stationary solution to the diffusion equation with absorbing boundary conditions for the initial conditions in \textbf{a)}. The stationary solution \textbf{b)} is the target for the NN. We fixed the diffusion constant to $D = 1\;voxels^2/s$ and the decay rate to $\gamma = 1/400s^{-1}$, which yields a diffusion length equal to $\sqrt{D/\gamma}\;voxels = 20 voxels$.
% \textcolor{red}{I think this should appear earlier, perhaps before figure 2 and maybe even before figure 1.}
% } \label{fig:inout0}
% \end{figure}

\begin{figure}[hbtp]
\centering
\subfigure[Training Loss without roll-back]{
\includegraphics[width=3.13in]{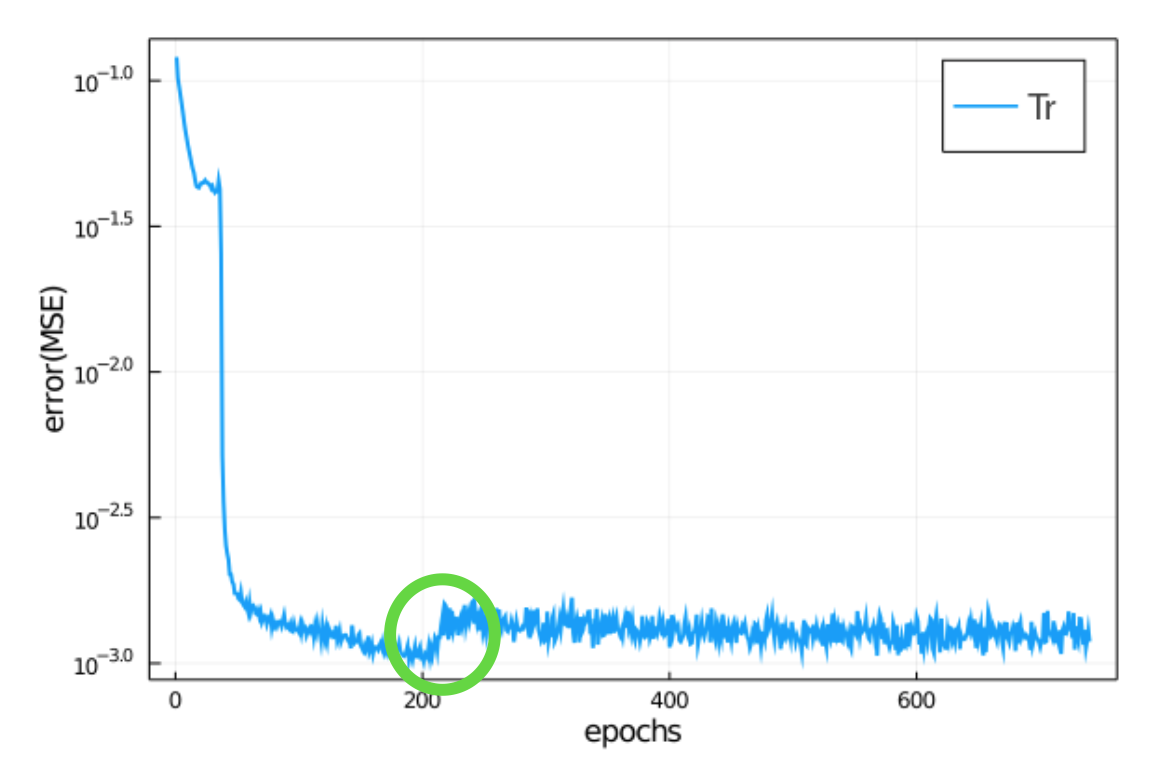}
\label{fig:withoutrollback}}
% \centering
% \subfigure[Test Loss Without roll back]{
% \includegraphics[width=3.1in]{Figs/TeWO.png}
% \label{fig:withoutrollbackTe}}
\centering
\subfigure[Training Loss with roll-back]{
\includegraphics[width=3.1in]{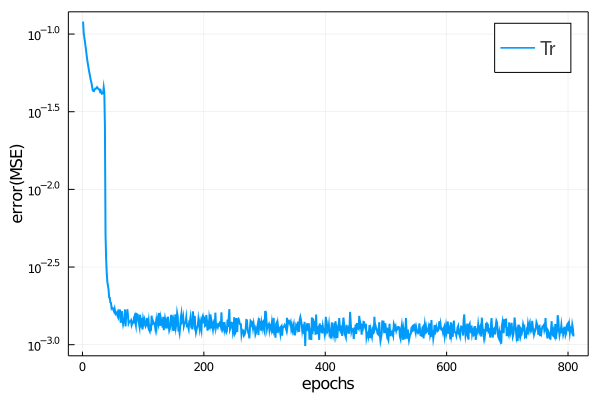}
\label{fig:withrollback}}
% \centering
% \subfigure[Test Loss With roll back]{
% \includegraphics[width=3.1in]{Figs/TeW.png}
% \label{fig:withrollbackTe}}
\caption{Training loss function \textit{vs} epochs for model 9 (the hyperparameters are specified in Table \ref{table:hyp} and the NN details are described in the main text) without roll-back \textbf{a)} and with roll-back \textbf{b)} using the same seed. We have circled in green where a jump occurred during this training run (see main text for discussion).
} \label{fig:rollback}
\end{figure}

We trained the CNN setting the number of epochs to $800$ using the deep learning library in Julia called Flux \cite{innes2018flux}. We varied the dropout values between $0.0$ and $0.6$ in steps of $0.1$ (see Table \ref{table:hyp}). We used ADAM as the optimizer \cite{kingma2014adam}. Deciding on a loss function is a critical choice in the creation of the surrogate. The loss function determines the types of error the surrogate's approximation will make compared to the direct calculation and the acceptability of these errors will depend on the specific application. The mean squared error (\textit{MSE}) error is a standard choice. However, it is more sensitive to larger absolute errors and therefore tolerates large relative errors at pixels with small values. A loss function calculated on the log of the values would be equally sensitive to relative error no matter what the absolute value. In most biological contexts we want to have a small absolute error for small values and a small relative error for large values. We explored the use of both functions, \textit{MAE} and \textit{MSE}, as described in Table \ref{table:hyp}. We used $80\%$ and $20\%$ of the dataset for training and test sets, respectively. We trained each model once. The highest and lowest values in the input and output images are $1$ and $0$, respectively. The former only occurs in sources and their vicinity. Given the configurations of the sources, the fraction of pixels in the image with values near $1$ is $\sim 2\pi R^2/L^2 \approx 2\%$. 
Thus, pixels with small values are much more common than pixels with large values, and because the loss function is an average over the field, high field values tend to get washed out.
To account for this unbalance between the frequency of occurrence of low and high values, we introduced an exponential weight on the pixels in the loss function. We modulate this exponential weight through a scalar hyperparameter $w$, for the field in the $i$th lattice position in the loss function as 
% jps: the equation below uses the boldmath "\bm" package  
\begin{equation}
\mathcal{L}_{i \beta}^{(\alpha)} = \exp(-(\langle i |\bm{1}\rangle- \langle i | y_{\beta} \rangle)/w) \cdot \left(\langle i | \hat{y}_{\beta} \rangle - \langle i | y_{\beta} \rangle \right)^\alpha  \; , \label{eq:Loss_weight}
\end{equation}
where $\alpha$ is $1$ or $2$ for MAE or MSE, respectively and $\beta$ tags the tuple in the data set (input and target). Here $\langle | \rangle$ denotes the inner product and $|i\rangle$ is a unitary vector with the same size as $|y_{\beta} \rangle$ with all components equal to zero except the element in position $i$ which is equal to one. $|\bm{1} \rangle$ is a vector with all components equal to 1 and with size equal to that of $|y_{\beta} \rangle$. Then $\langle i | y_{\beta} \rangle$ is a scalar corresponding to the pixel value at the $i$th position in $|y_{\beta} \rangle$, whereas $\langle i | \bm{1}\rangle=1$ for all $i$. Notice that high pixel values will then have an exponential weight $\approx 1$ while low pixel values will have an exponential weight $\approx \exp(-1/w)$. This implies that the error associated to high pixels will have a larger value than that for low pixels. The loss function $\mathcal{L}^{(\alpha)}$ is the mean value over all pixels ($i$) and a given data set ($\beta$):
\begin{equation}
\mathcal{L}^{(\alpha)} = \langle \mathcal{L}_{i \beta}^{(\alpha)} \rangle \; ,
\end{equation}
where $\langle \rangle$ denotes average. In our initial trial training runs, we noticed that the loss function always reached a plateau by $800$ epochs, so we trained the NNs over $800$ epochs for all runs reported in this paper. Because the training is stochastic, the loss function can increase as well as decrease between epochs as seen in Fig \ref{fig:rollback}. At the end of 800 epochs we adopted the network configuration with the lowest loss function regardless of the epoch at which it was achieved. 
%For each training series we kept the model with the lowest loss function value. 

While the trendline (averaged over 5 or ten epochs) of the loss function value tends to decrease during training, the stochasticity of the training means that the value of the loss function often increases significantly between successive epochs, even by one or two orders of magnitude (see Fig. \ref{fig:rollback}). In some cases, the loss function decreases back to its trend after one or two epochs, in other cases (which we call \textbf{jumps}), it stays at the higher value, resetting the trend line to the higher value and only gradually begins to decrease afterwards. In this case all of the epochs after the jump have larger loss functions than the epoch immediately before the jump, as shown for the evolution of the loss function for a typical training run in Fig. \ref{fig:withoutrollback}. This behavior indicates that the stochastic optimization algorithm has pursued an unfavorable branch. To avoid this problem, we added a \textit{roll-back} algorithm to the training, as proposed in Ref. \cite{Fox2020}. We set a loss threshold value, $\mathcal{L}_{thrs}$, such that if the ratio of loss value from epoch $n$ to $n+1$ is larger than $\mathcal{L}_{thrs}$, then the training algorithm reverts (\textbf{rolls back}) to the NN state corresponding to epoch $n-s$ and tries again. The stochasticity of training means that  roll-back has an effect similar to training an ensemble of models with the same hyperparameters and selecting the model with the lowest loss function value, however, the roll-back optimization takes much less computer time than a large ensemble. We set $s = 5$ and set the threshold value $\mathcal{L}_{thrs}$ to
\begin{equation}
    \mathcal{L}_{thrs} = C \frac{1}{m} \sum_{ep=n-m+1}^n \mathcal{L}^{(\alpha)}(ep) \; .
\end{equation}
Here we chose $C=5$ and $m=20$ where $ep$ stands for epoch, \textit{i.e.}, we set the threshold value to 5 times the average loss function value over the previous $m=20$ epochs. We chose these values empirically. In Fig. \ref{fig:withrollback} we have plotted a typical example of the evolution of the loss function during training when we train using roll-back. A typical number of roll-backs is $40$, i.e., this number is the number of epochs where the jump was higher than the threshold during the training.

% \textcolor{red}{jps: we never discuss figure 4. I think figure 4 and 5 should swapped and we should discuss figure 4 at the start of Results and Discussion. Should emphasise  that qualitatively the NN is getting the output pattern correct.}

%\section{Results and Discussion}
\section{Results}

% \textcolor{red}{jps: separated "Results and Discussion" into separate sections.}

% \textcolor{red}{jps: start with discussing the figure 4 here saying that qualitatively the results on test sets (that's what shown in figure 4?) are qualitative very good. Then say which NN layout was best.}

\begin{figure}[hbtp]
\centering
\subfigure[Input]{
\includegraphics[width=3.1in, height=1.9in]{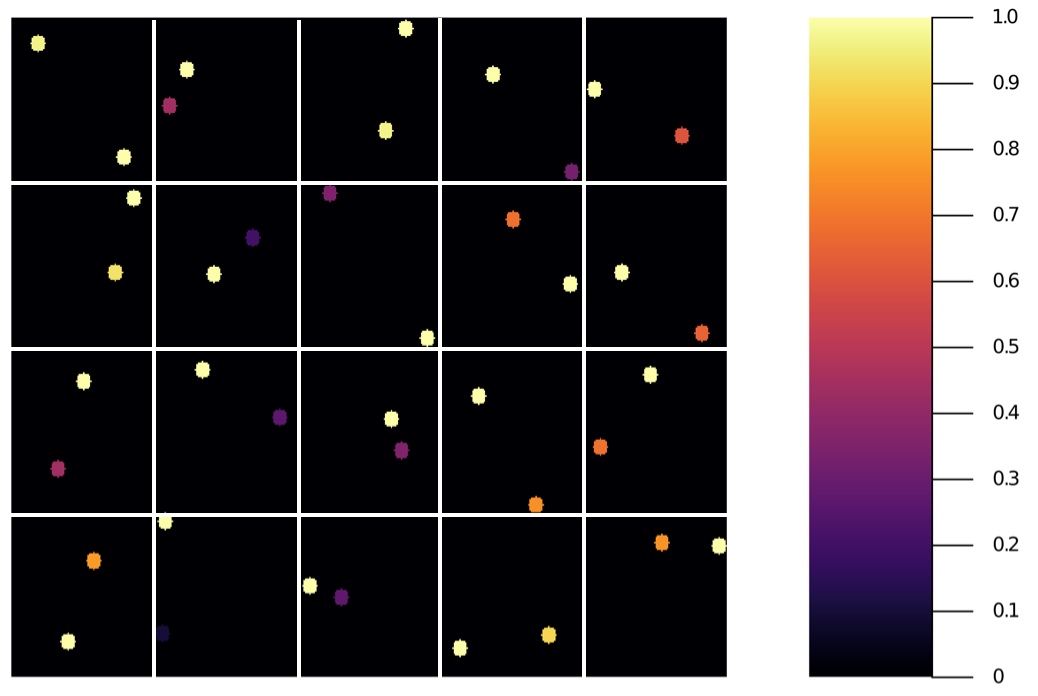}
\label{fig:input}}
\centering
\subfigure[Ground truth]{
\includegraphics[width=3.1in, height=1.9in]{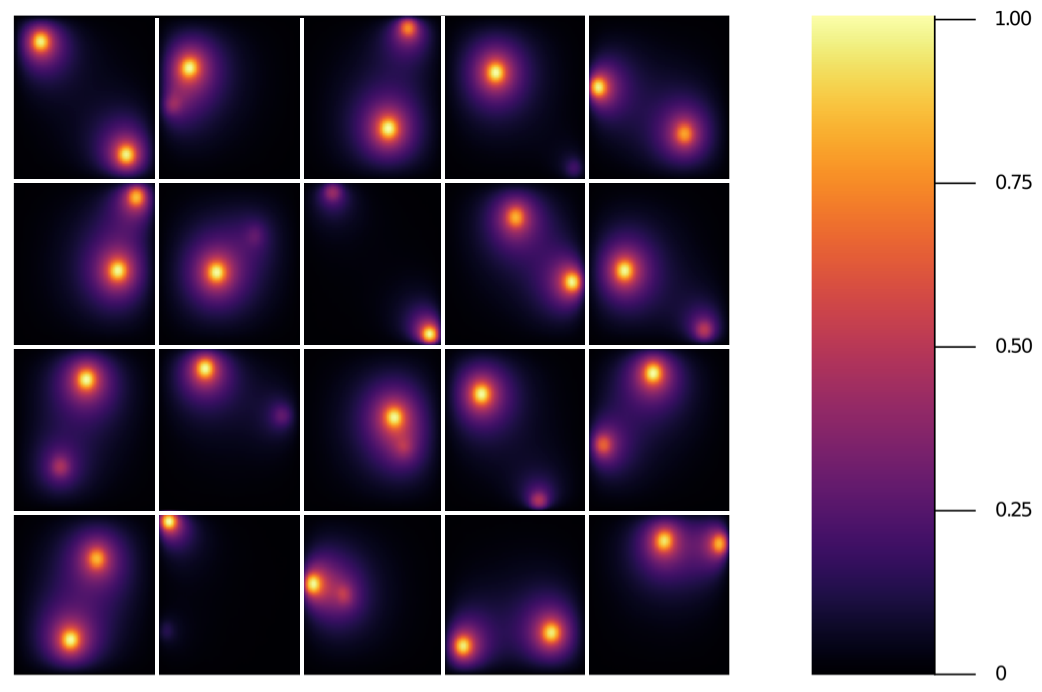}
\label{fig:target}}
\centering
\subfigure[Prediction]{
\includegraphics[width=3.1in, height=1.9in]{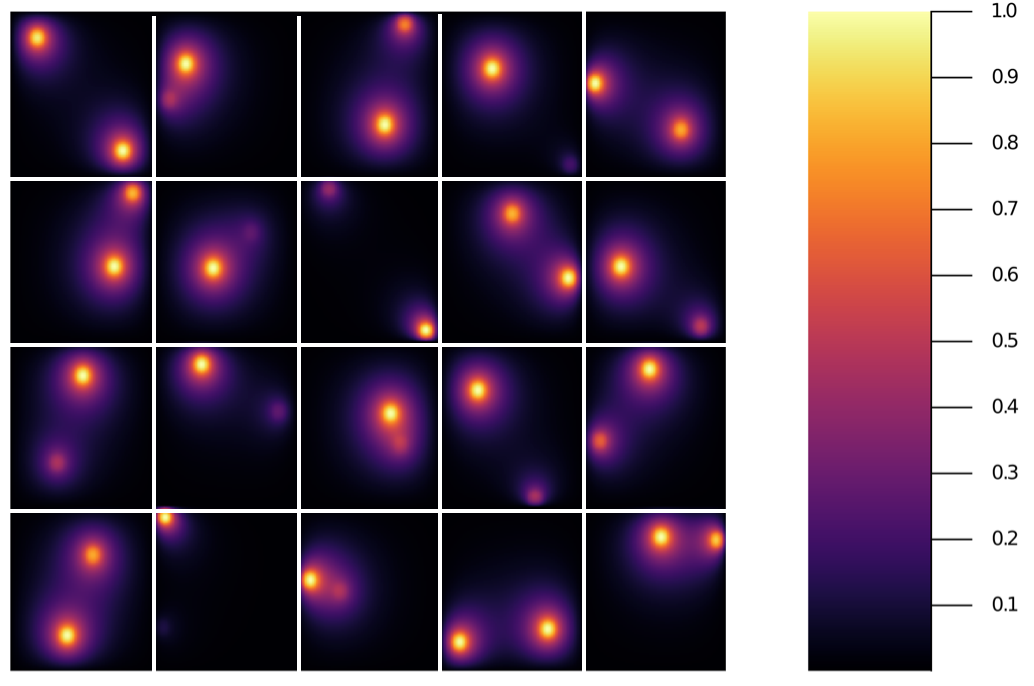}
\label{fig:output}}
% \centering
% \subfigure[Without roll back]{
% \includegraphics[width=3.1in]{Figs/withoutrollback.png}
% \label{fig:withoutrollback}}
% \centering
% \subfigure[With roll back]{
% \includegraphics[width=3.1in]{Figs/withrollback.png}
% \label{fig:withrollback}}
\caption{ Results for 20 randomly selected test data sets' \textbf{a)} input, \textbf{b)} ground truth (target output) and  \textbf{c)} NN surrogate prediction of steady-state diffusion field output for the input. %\textcolor{red}{jps: Are these training or test data sets? Is this 20 random training sets selected from 4K total test sets, or are these all from the 16k of training sets?  \textbf{\underline{We don't refer to this figure (and this figure doesn't have a label?) in the text!}} Probably should after discussing "rollback". Which parameter set is this for? Why does the Input panel have colors and a color bar? The values for the sources are random (0,1) but the panels seem to be pretty highly biased to high values.} 
%\textbf{a)} We placed two random value sources of radius $5\;voxels$ in random positions fully within a $100\times100 pixel$ lattice and used this configuration as the input to the NN. \textbf{b)} Stationary solution to the diffusion equation with absorbing boundary conditions for each of the initial conditions in \textbf{a)}. 
% jps: The stationary solution \textbf{b)} is the target for the NN. \textbf{c)} NN surrograte predictions corresponding to the inputs $\textbf{a)}$. 
%We fixed the diffusion constant to $D = 1\;voxels^2/s$ and the decay rate to $\gamma = 1/400s^{-1}$, which yields a diffusion length equal to $\sqrt{D/\gamma}\;voxels = 20 voxels$. Training set (blue) and test set (orange) loss \textit{vs} epochs for model 9 without roll-back \textbf{d)} and with roll-back \textbf{e)} using the same seed. We have circled in green where the jump happens.
} \label{fig:inout}
\end{figure}

\begin{figure}[hbtp]
\centering
\subfigure[Ground truth]{
\includegraphics[width=3.1in, height=1.9in]{Figs/target6.png}
\label{fig:target2}}
\centering
\subfigure[Snapshot of the Residual in a batch]{
\includegraphics[width=3.1in, height=1.9in]{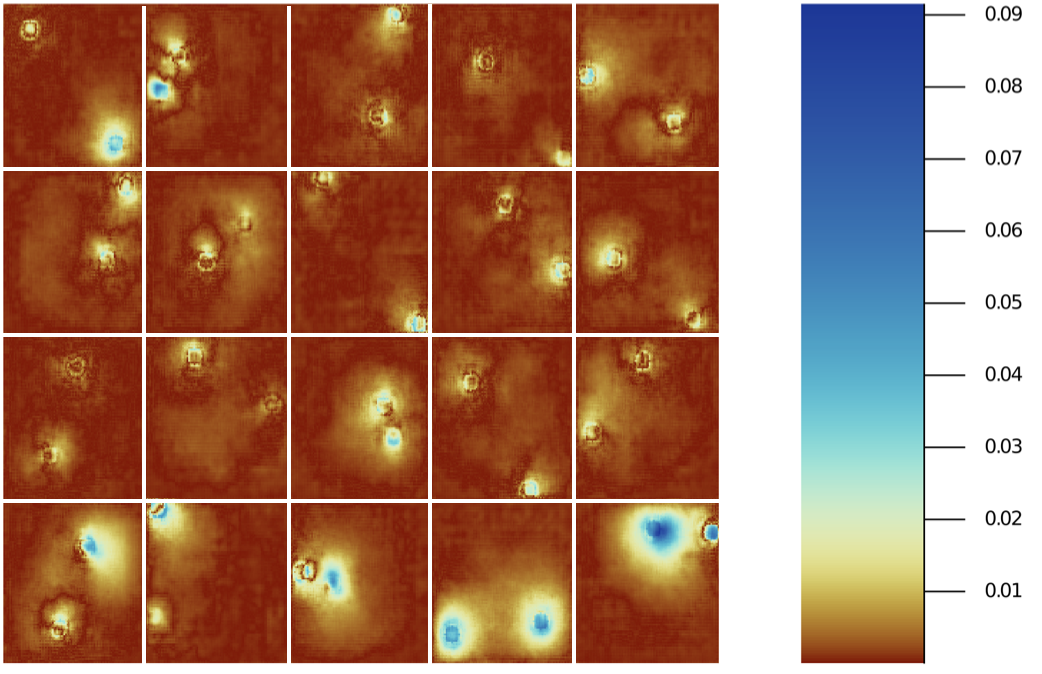}
\label{fig:res}}
\centering
\subfigure[Snapshot of the Residual/true value  in a batch]{
\includegraphics[width=3.1in, height=1.9in]{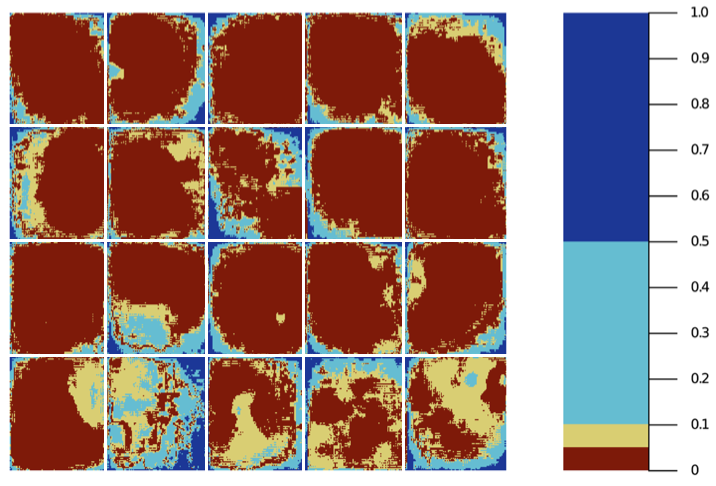}
\label{fig:chi}}
\centering
\subfigure[Mean value, 99-Percentile and max residual \textit{vs} models]{
\includegraphics[width=3.1in]{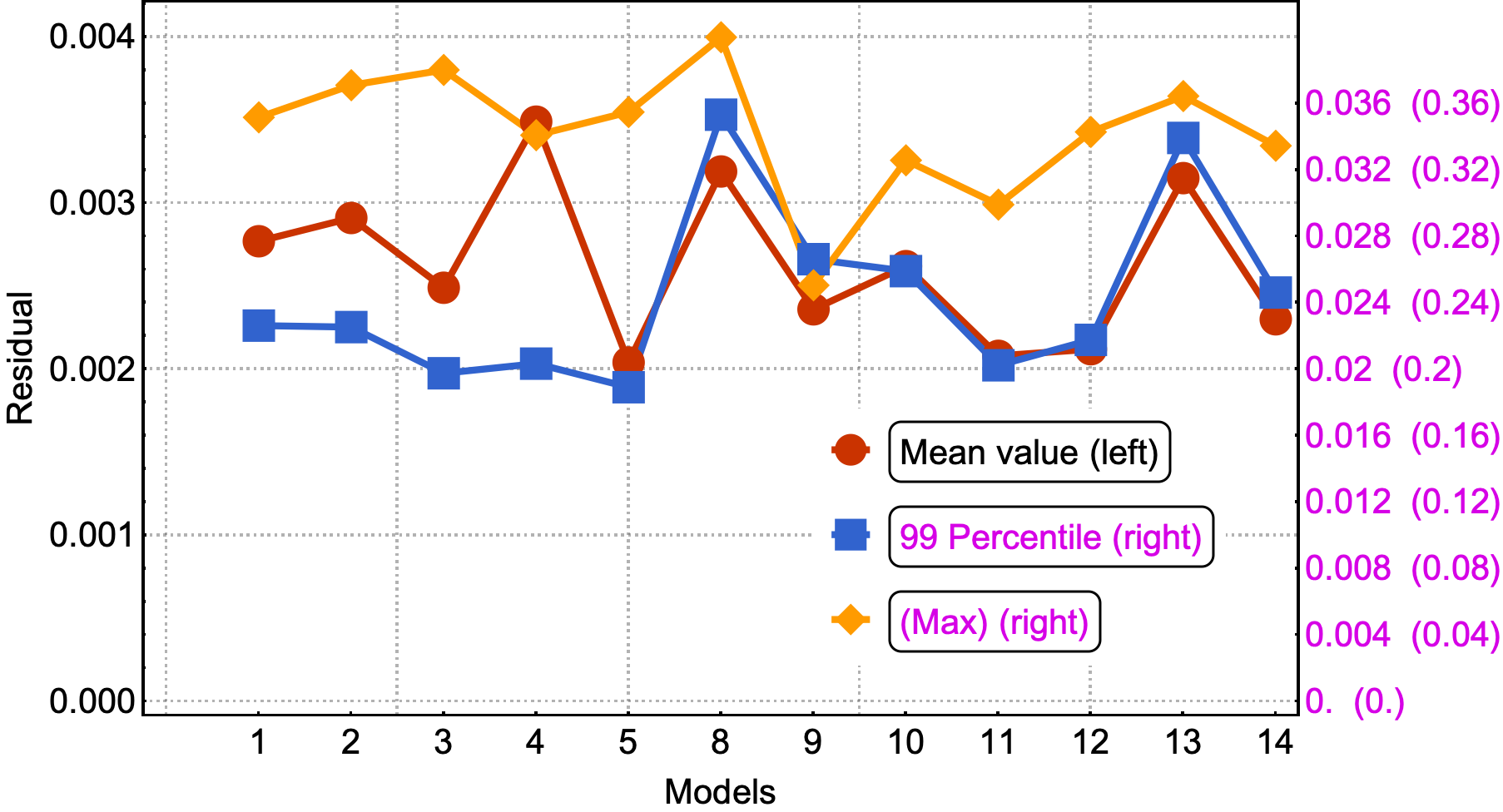}
\label{fig:res_error}}
\caption{\textbf{a)} The stationary solution for the same batch in the test set. \textbf{b)} Residual (absolute error, i.e., $| |y_{\beta} \rangle - | \hat{y}_{\beta} \rangle|$) for twenty sample source images in the test set trained using model 12 in Table \ref{table:hyp}. \textbf{c)} Residual/true value (relative error) for the corresponding images. \textbf{d)} Mean, 99-Percentile and maximum residual for all of the models in Table \ref{table:hyp}. Left scale for mean value, right scale for 99-Percentile residual value and right scale in parentheses for max residual value.} \label{fig:comparison}
\end{figure}

Quite commonly, the mean residual is the estimator used to judge the goodness of a given model. However, there are cases where the worst predictions are highly informative and can be used to make basic decisions about which features of the NN do not add value. In Figs. \ref{fig:input}, \ref{fig:target} and \ref{fig:output} we show 20 different inputs, targets and predictions, respectively. 
The predictions in Fig. \ref{fig:output} were obtained using model 12 (see Table \ref{table:hyp}) and qualitatively show very good results. 
For each model we computed the residual, \textit{i.e.}, the absolute value of the difference between the ground truth and the NN prediction pixel-by-pixel, as shown in Fig. \ref{fig:res}. We also analyzed the relative residual, \textit{i.e.}, the residual divided by the ground truth pixel-by-pixel, as shown in Fig \ref{fig:chi}. Models 6 and 7, which only use NN 1 ($p_1=1$ and $p_2=0$), yield mean residuals an order of magnitude larger than models that use both or only NN 2. Therefore, we  reject the NN 1-only models and do not analyze them further. 

Table \ref{table:hyp} summarizes the hyperparameter values for each model we trained. The choice of these parameters was empirically driven. Since we had the field values bounded between $0$ and $1$ similar to black and white images, we tested different $L$-norms, namely, mean absolute value (MAE), mean squared value (MSE) and mean to the fourth power, often used in neural networks applied to images. In this paper we show the results for MAE and MSE. We also tested different hyperparameters values for the dropout. We found that low dropout values for NN 2 yield the best results. 
%In addition, we trained models using momentum as the optimizer as well as models using the mean residual to the fourth power as a loss function. The results for those models were qualitatively worse. Therefore, we don't show them since it would cloud the results obtained from the models described in Table \ref{table:hyp}. 

%[Need to say that the model we pick as the "best" model will depend on the metrics we use to evaluate quality and that we will consider these issues for single metrics and for ensembles of metrics.... 
%Be specific in distinguishing cases where you only used NN1, where you only used NN2 and where you used a mixture. Explain what you found in each case....]

%Besides, as a matter of good practice, worst case predictions should always be included as part of the evaluation of any NN. 

In Fig. \ref{fig:res_error} we have plotted the mean residual value, the 99-Percentile residual value and the maximum residual value computed over the test set. 
Notice that the 99-Percentile residual value is ten times the mean residual value and the maximum residual value is ten times the 99-Percentile residual value. This suggests that the residual distribution contains outliers, i.e., there is a $1\%$ residual that deviate from mean residual 10 to 100 times. Furthermore, these outliers correspond to regions between the source and the border,  near the source, where the source is close to the border as suggested by Fig. \ref{fig:res}. While the largest values in absolute residual come from pixels near the source as shown in Fig. \ref{fig:res}, the relative error near the source is small whereas the relative error near boundaries is large, as shown in Fig. \ref{fig:chi}. Since we are considering absorbing boundary conditions, the field at the boundary is always equal to zero, thus strictly speaking the relative residual value has a singularity at the boundary. Thus, at the boundaries there is a larger relative error due to the boundary conditions.

Models 5, 11 and 12 have low mean residuals with model 5 being the smallest. Focusing instead on the mean residual and the 99-Percentile, we notice that models 3, 4, 5, 11 and 12 yield the best results. Finally, considering the maximum residual together with the previous estimators, we notice that model 9 has low mean residual, low 99-percentile residual and the lowest max residual. Depending on the user's needs, one estimator will be more relevant than others. In this sense, defining a \textit{best} model is relative. Nevertheless, having more metrics (e.g. relative error for large values and absolute error for small values) helps to characterize each model's performance. In future work we'll consider more adaptable metrics, as well as mixed error functions that incorporate multiple estimators.
%like relative error for large values and absolute error for small values

\begin{figure}[hbtp]
\centering
\subfigure[Model 9 Test]{
\includegraphics[width=2.8in]{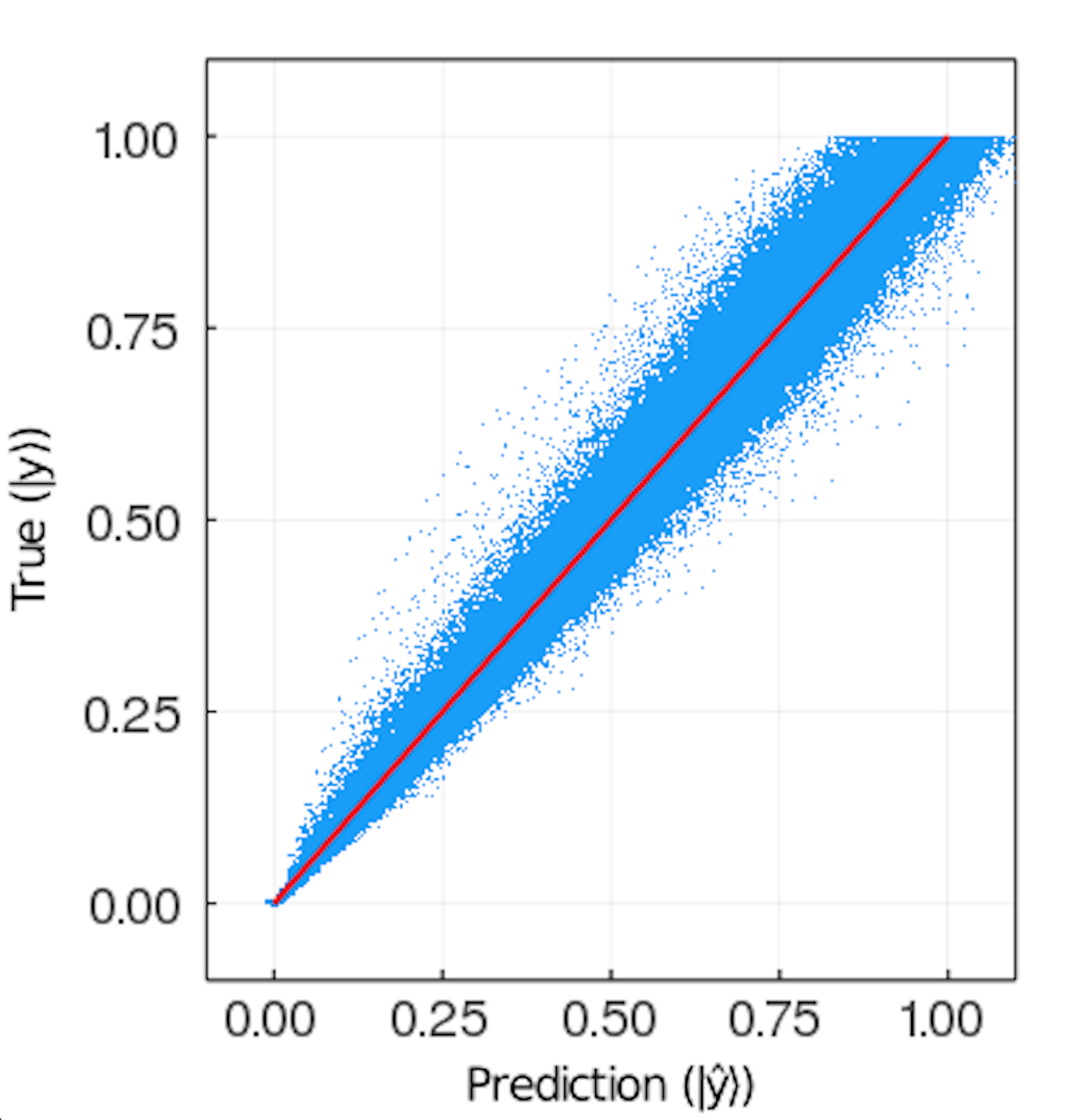}
\label{fig:11-test}}
\centering
\subfigure[Model 9 Training]{
\includegraphics[width=2.8in]{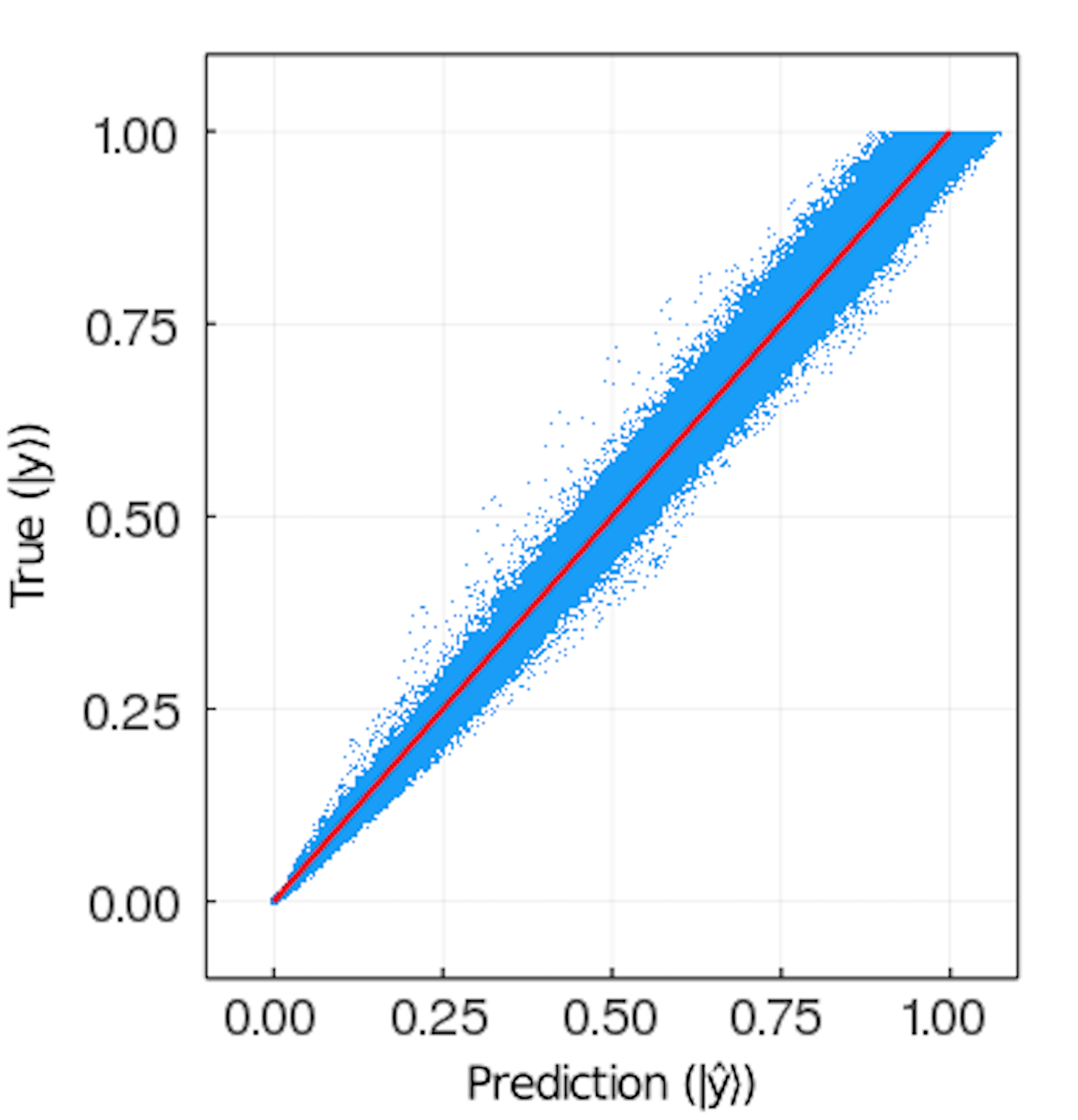}
\label{fig:11-training}}
\centering
\subfigure[Model 11 Test]{
\includegraphics[width=2.8in]{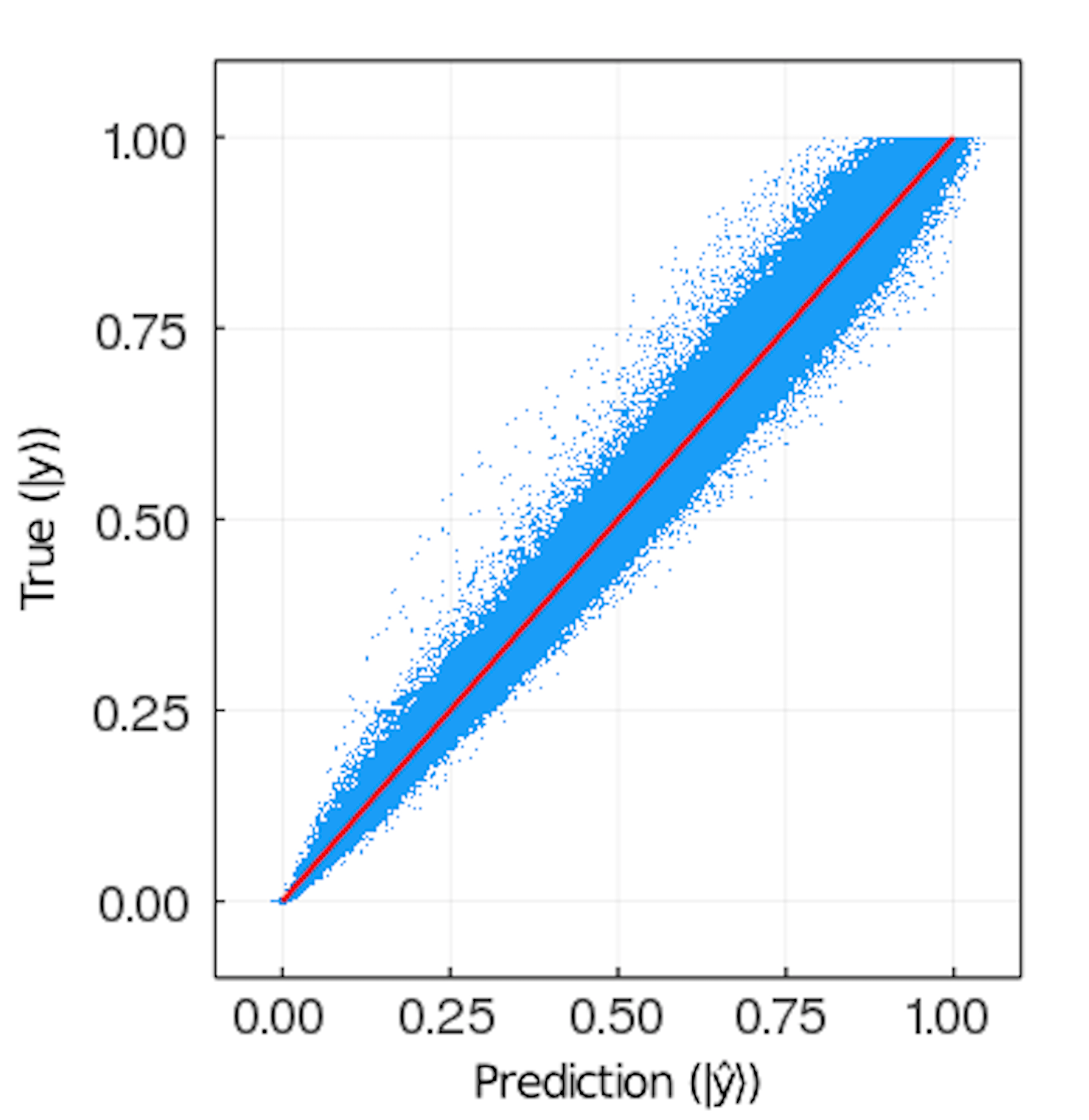}
\label{fig:9-test}}
\centering
\subfigure[Model 11 Training]{
\includegraphics[width=2.8in]{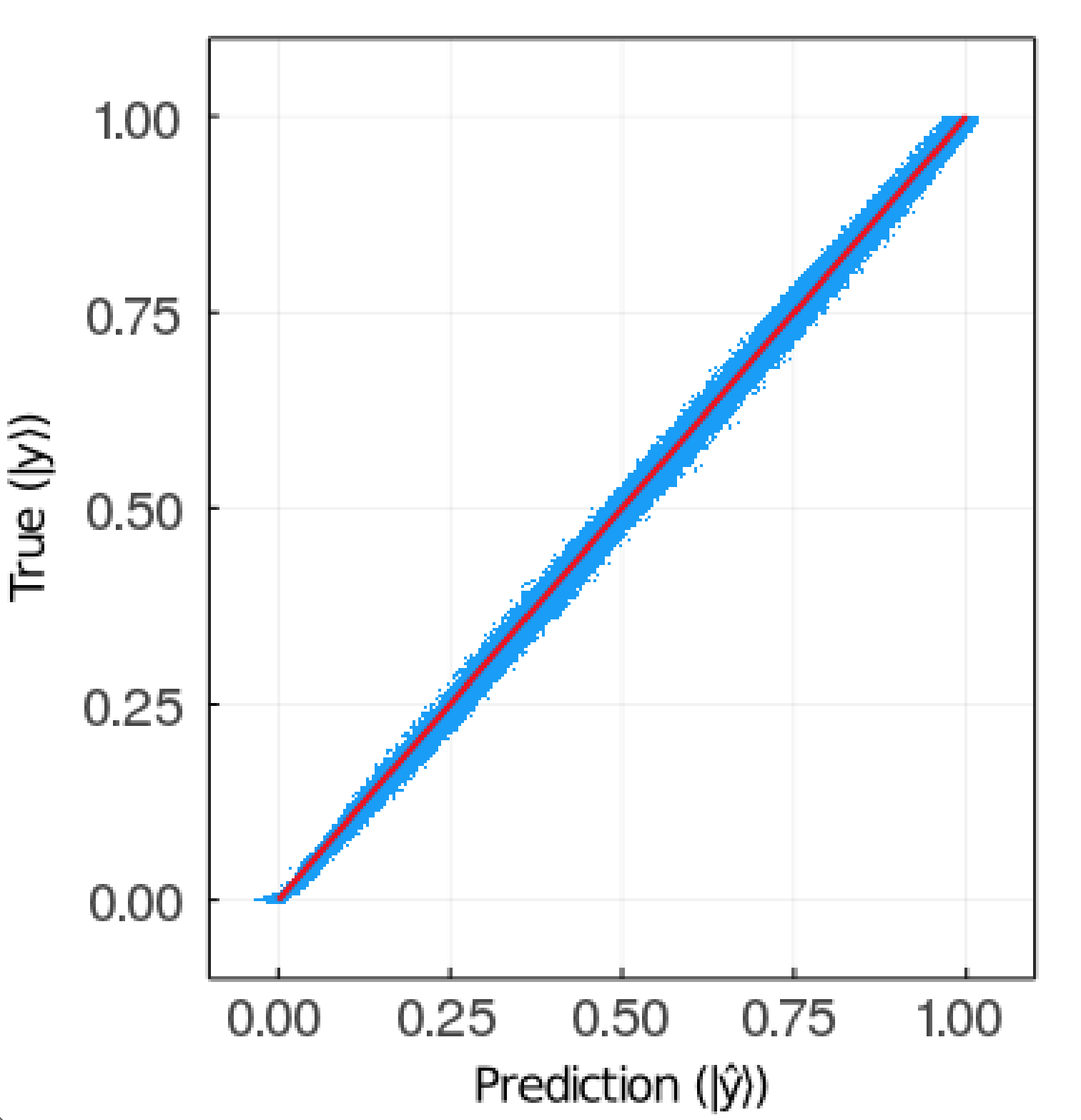}
\label{fig:9-training}}
\caption{Ground truth \textit{vs} prediction for \textbf{a)} test set and \textbf{b)} training set in the case of model 9; \textbf{c)} test set and \textbf{d)} training set in the case of model 11  (see Table \ref{table:hyp}). The number of points plotted in each panel is $3.75\cdot 10^7$. } \label{fig:true-output}
\end{figure}

Fig. \ref{fig:true-output} plots the prediction \textit{versus} the target for each pixel in each image in the training and test sets for models 9 and 11. Notice that for the test sets the results are qualitatively similar between models, for the training set the dispersion is larger in model 11 than in model 9. This suggests model 11 is overfitting the training data. Models 9 and 11 have the same hyperparameters except for the weight $w$. In the former $w=100$ while in the latter $w=1$. This suggests that the exponential weight helps reduce overfitting. \label{par:1}

\begin{table}[h!]
\centering
{\renewcommand{\arraystretch}{1.3} % jps  add a bit of extra space above rows
 \begin{tabular}{|c c c c c c c c c|} 
 \hline
 Model & weight (w) & $p_1$ & $p_2$ & $D_1 \; D_2 \; D_3 \; D_4$  & Loss & $\langle$res$\rangle$ ($10^{-3}$) & 99-P res ($10^{-2}$) & max res \\ [0.5ex] 
 \hline\hline\hline
 1 & 1000 & 1 & 1 & $0.3 \; 0.3 \; 0.3 \; 0.3$ & MSE & 2.77 & 2.26 & 0.35 \\
 \hline
 2 & 1 & 1 & 1 & $0.3 \; 0.3 \; 0.3 \; 0.3$ & MSE & 2.91 & 2.25 & 0.37 \\ 
 \hline
 3 & 1 & 1 & 1 & $0.4 \; 0.4 \; 0.1 \; 0.1$ & MSE & 3.49 & 2.03 & 0.34 \\ 
 \hline
 4 & 1 & 0 & 1 & $- \; - \; 0.3 \; 0.3$ & MSE & 2.49 & 1.97 & 0.38 \\
 \hline
 5 & 1 & 0 & 1 & $- \; - \; 0.1 \; 0.1$ & MSE & 2.04 & 1.89 & 0.35 \\
 \hline
 6 & 1 & 1 & 0 & $0.3 \; 0.3 \; - \; -$ & MSE & 75.8 & 16.5 & 0.47 \\
 \hline
 7 & 1 & 1 & 0 & $0.4 \; 0.4 \; - \; -$ & MSE & 79.9 & 21.6 & 0.65 \\
 \hline\hline
 8 & 100 & 1 & 1 & $0.3 \; 0.3 \; 0.3 \; 0.3$ & MAE & 2.62 & 2.59 & 0.33 \\
 \hline
 9 & 100 & 1 & 1 & $0.4 \; 0.4 \; 0.1 \; 0.1$ & MAE & 2.08 & 2.02 & 0.30 \\ 
 \hline
 10 & 1 & 1 & 1 & $0.3 \; 0.3 \; 0.3 \; 0.3$ & MAE & 3.19 & 3.53 & 0.40 \\ 
 \hline
 11 & 1 & 1 & 1 & $0.4 \; 0.4 \; 0.1 \; 0.1$ & MAE & 2.36 & 2.66 & 0.25 \\ 
 \hline
 12 & 1 & 0 & 1 & $- \; - \; 0.1 \; 0.1$ & MAE & 2.12 & 2.17 & 0.34 \\ 
 \hline
 13 & 10 & 0 & 1 & $- \; - \; 0.3 \; 0.3$ & MAE & 3.15 & 3.39 & 0.36 \\ 
 \hline
 14 & 10 & 0 & 1 & $- \; - \; 0.1 \; 0.1$ & MAE & 2.30 & 2.46 & 0.33 \\ 
 \hline
 %15 & 1000 & $(0.3,0.3,0.3, 0.3)$ & 1 & 1 & MSE & ADAM \\ 
 %\hline
\end{tabular}
} %%% jps  end the extra vertical space in the table
\caption{Trained models with their corresponding hyperparameters. Each model is numbered for reference. The weight $w$ is defined in Eq. \eqref{eq:Loss_weight}. The $D_i$ for $i=1,...,4$ are the dropout values (see Table. \ref{table:1}). $D_1$ and $D_2$ apply to NN 1 whereas $D_3$ and $D_4$ apply to NN 2. $p_1$ and $p_2$ are Boolean variables. $p_i=0$ ($p_i=1$) implies NN $i$ is turned off (on). If $p_1 =0$ then the values of $D_1$ and $D_2$ are irrelevant, while $p_2 =0$ makes the values of $D_3$ and $D_4$ irrelevant. The loss column specifies the loss function, either \textit{MSE} for mean squared error ($\alpha = 2$) or mean absolute error \textit{MAE} ($\alpha = 1$), respectively (see Eq. \eqref{eq:Loss_weight}). The mean res, 99-P res and max res columns show the mean, 99-percentile and maximum residual for each model computed over the test set.}
\label{table:hyp}
\end{table}

\begin{figure}[hbtp]
\centering
\subfigure[Prediction from NN 1]{
\includegraphics[width=3.1in, height=1.9in]{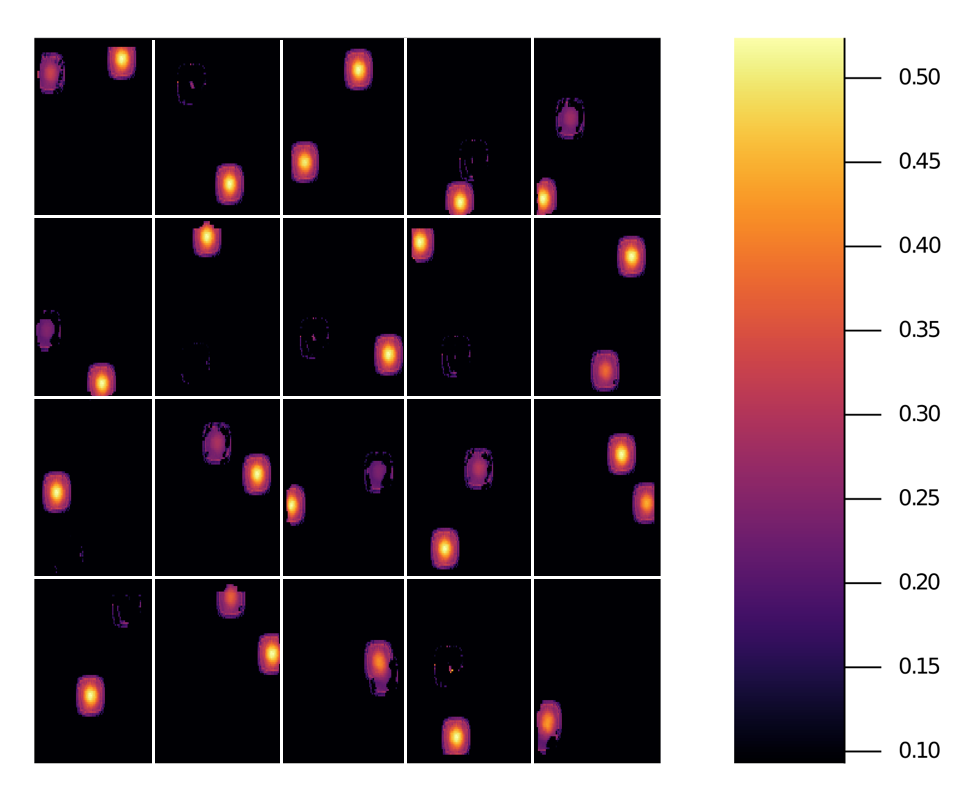}
\label{fig:n1}}
\centering
\subfigure[Prediction from NN 2]{
\includegraphics[width=3.1in, height=1.9in]{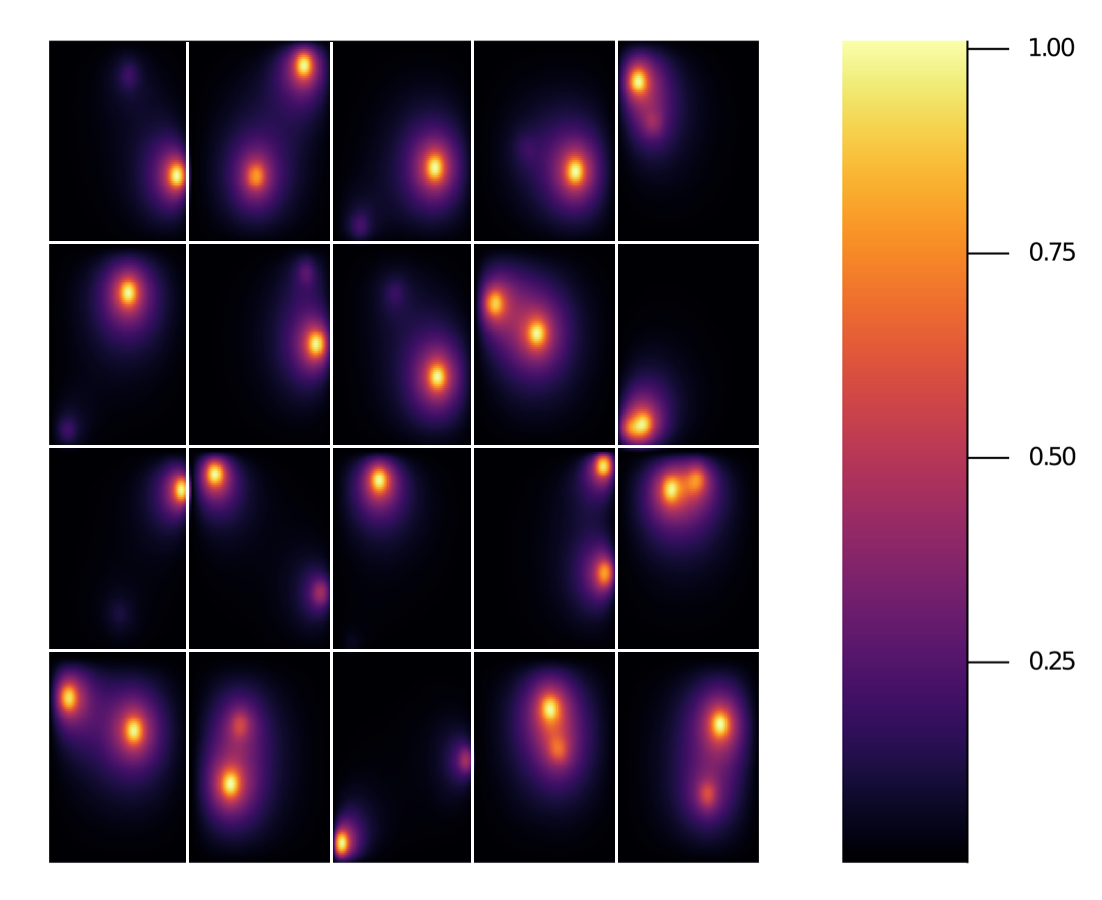}
\label{fig:n2}}
\caption{Results for 20 randomly selected test data sets \textbf{a)} Prediction using model 7, which only uses NN 1. \textbf{b)} Prediction using model 5, which only uses NN 2. See Table \ref{table:hyp}. Note the different scale on the color bars.
%, should we mention that in the caption?. So the NN1 model, with these parameters, never gives a maximum field value of more than about 0.5? The blockyness of NN1 compared to NN2 is also interesting.
} \label{fig:n1n2}
\end{figure}

%\textbf{You need to add a figure and a discussion which shows the output of NN1 and the output of NN2 separately and discusses what they do! At the moment you mention this repeatedly in the conclusion but have nothing about it in the main text}

\begin{figure}[hbtp]
\centering
% \subfigure[Model 11 Test]{
\includegraphics[width=6.4in]{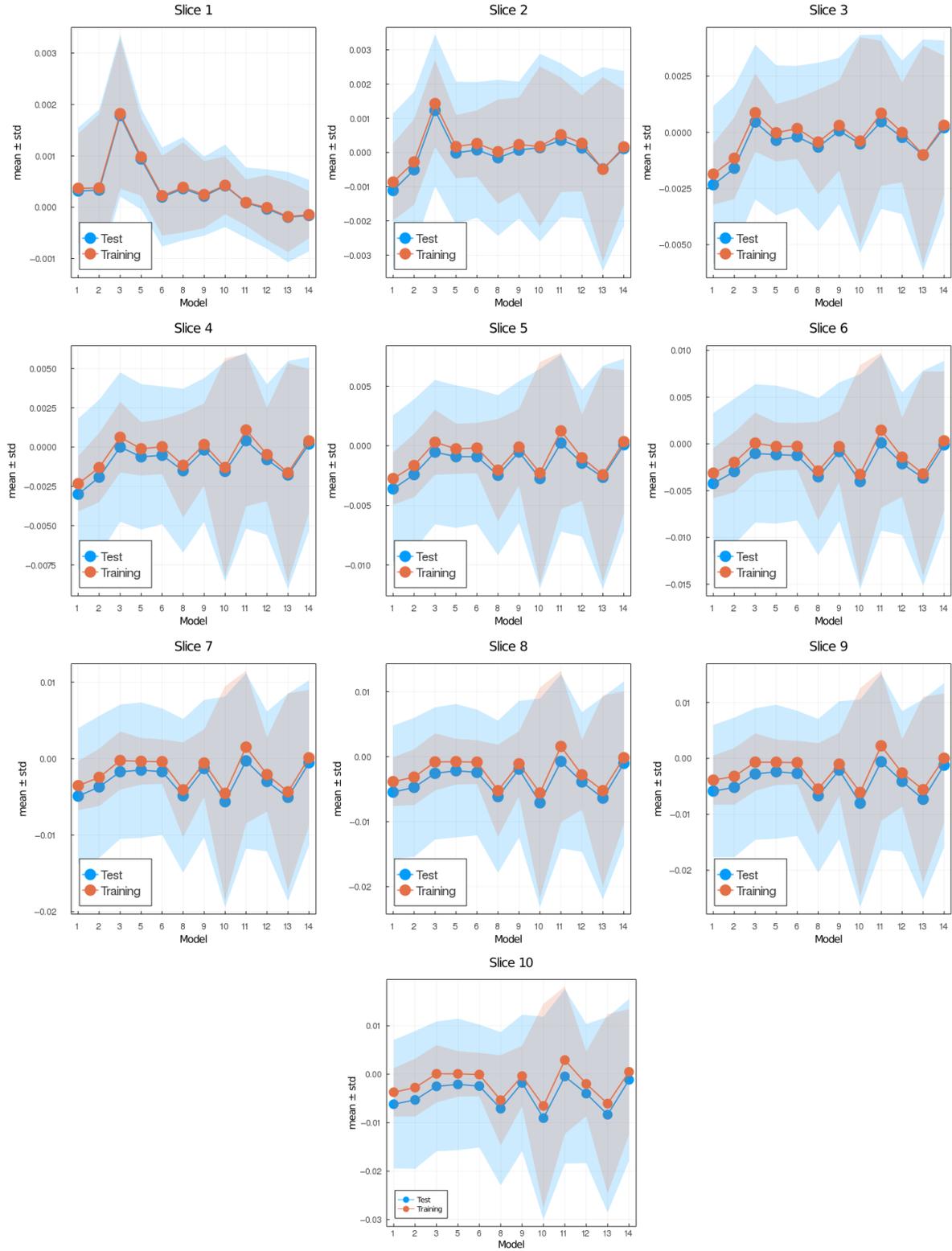}
% \label{fig:11-test}}
\caption{Mean (data points) $\pm$ standard deviation (envelop) per slice \textit{vs} models (see Table \ref{table:1}) for test set (blue) and training set (red). Slice $i$ corresponds to field values in the interval $[0.1\cdot (i-1), 0.1 \cdot i ]$ where $i=1,...,10$.} \label{fig:Slice-plot}
\end{figure}

In Fig. \ref{fig:n1n2} we show the prediction from NN 1 (Fig. \ref{fig:n1}) and NN 2 (Fig. \ref{fig:n2}). Notice that NN 1 is able to detect the sources whereas NN 2 is able to predict the field. Using both neural networks improves the results as can be seen in Fig. \ref{fig:res_error}.
As previously mentioned, pixels with low (near 0) field values are much more common than pixels with high (near 1) field values. While  the exponential factor in the loss function compensates for this bias, the residual in Fig. \ref{fig:res_error} does not. To address this issue we compute the mean residual over small field intervals. This will tell us how well the model predicts for each range of absolute values. Furthermore, this method can be used to emphasize accuracy or relative accuracy in different value ranges. The way we do this is as follows. In Fig. \ref{fig:true-output} we take $10$ slices of size $0.1$ in the direction $y=x$. We then compute the mean residual and standard deviation per slice. In section Supplement \ref{app:PDF_Slices} 
%\textcolor{red}{should this be \ref{fig:Plot_s} ? the figure reference is wrong I think. Or, is this supposed to be the supplement?}
we have plotted the PDF (probability density function) per slice (blue bins) and a Gaussian distribution (red curve) with mean and standard deviation set to the mean residual and standard deviation per slice, respectively. We did this for all models in Table \ref{table:hyp}. In Fig. \ref{fig:Slice-plot} we  plotted the mean residual \textit{vs} for each model for each slice for the test and training sets. The error envelop shows the residual standard deviation per slice. Notice that models trained with MSE have a smaller residual standard deviation than models trained with MAE in the case of the training set, which suggest that MSE contributes to overfitting more that MAE. Recall that the difference between the MSE gradient and the MAE gradient is that the former is linear with the residual value whereas the latter is a constant. Therefore, training with MAE generalizes better than MSE.  Additionally, notice the dispersion increases with the slice number.

\begin{figure}[hbtp]
\centering
\subfigure[Average of the mean residual over slices]{
\includegraphics[width=3.1in]{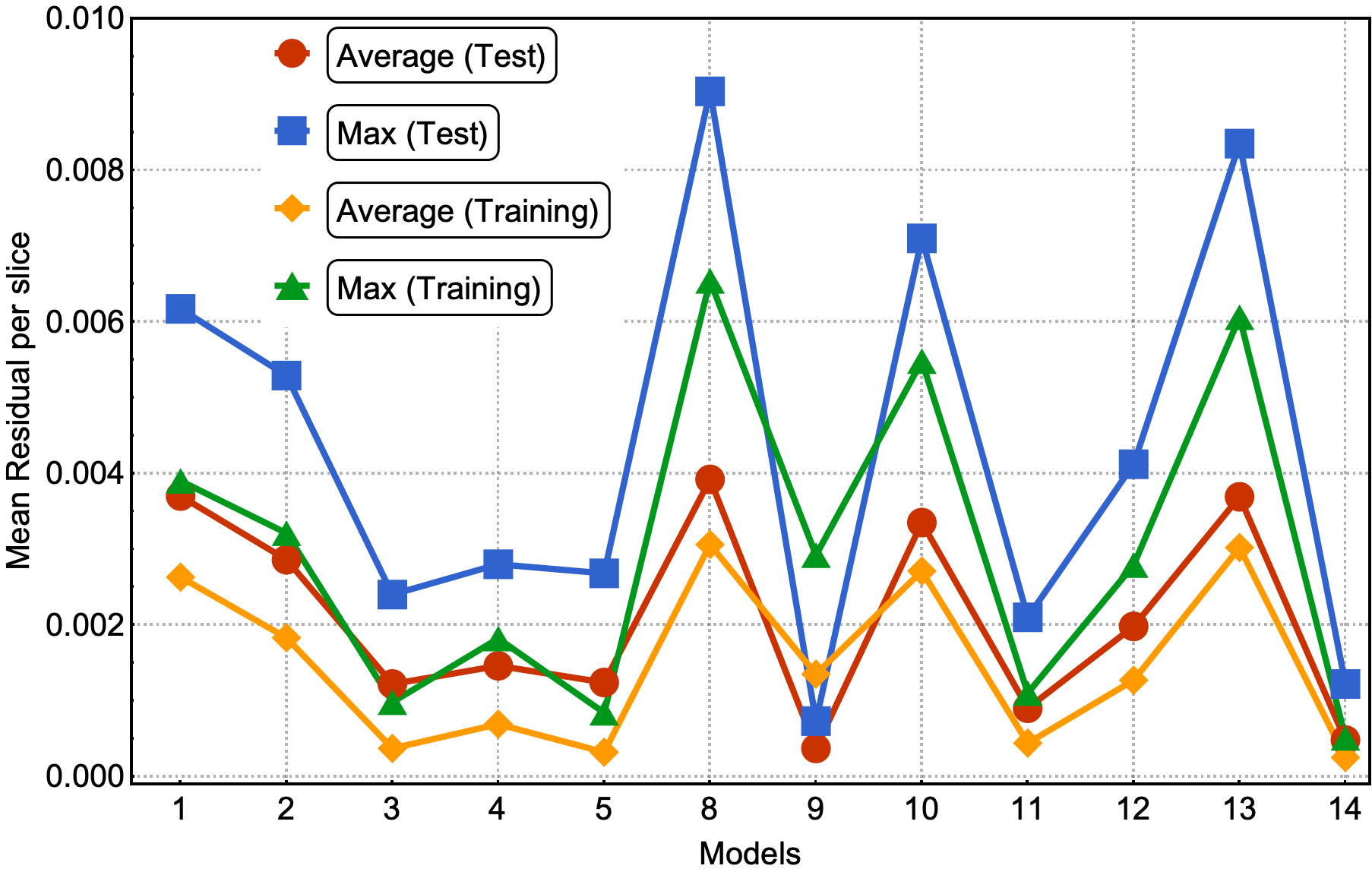}
\label{fig:Plot_m}}
\centering
\subfigure[Average of the residual standard deviation over slices]{
\includegraphics[width=3.1in]{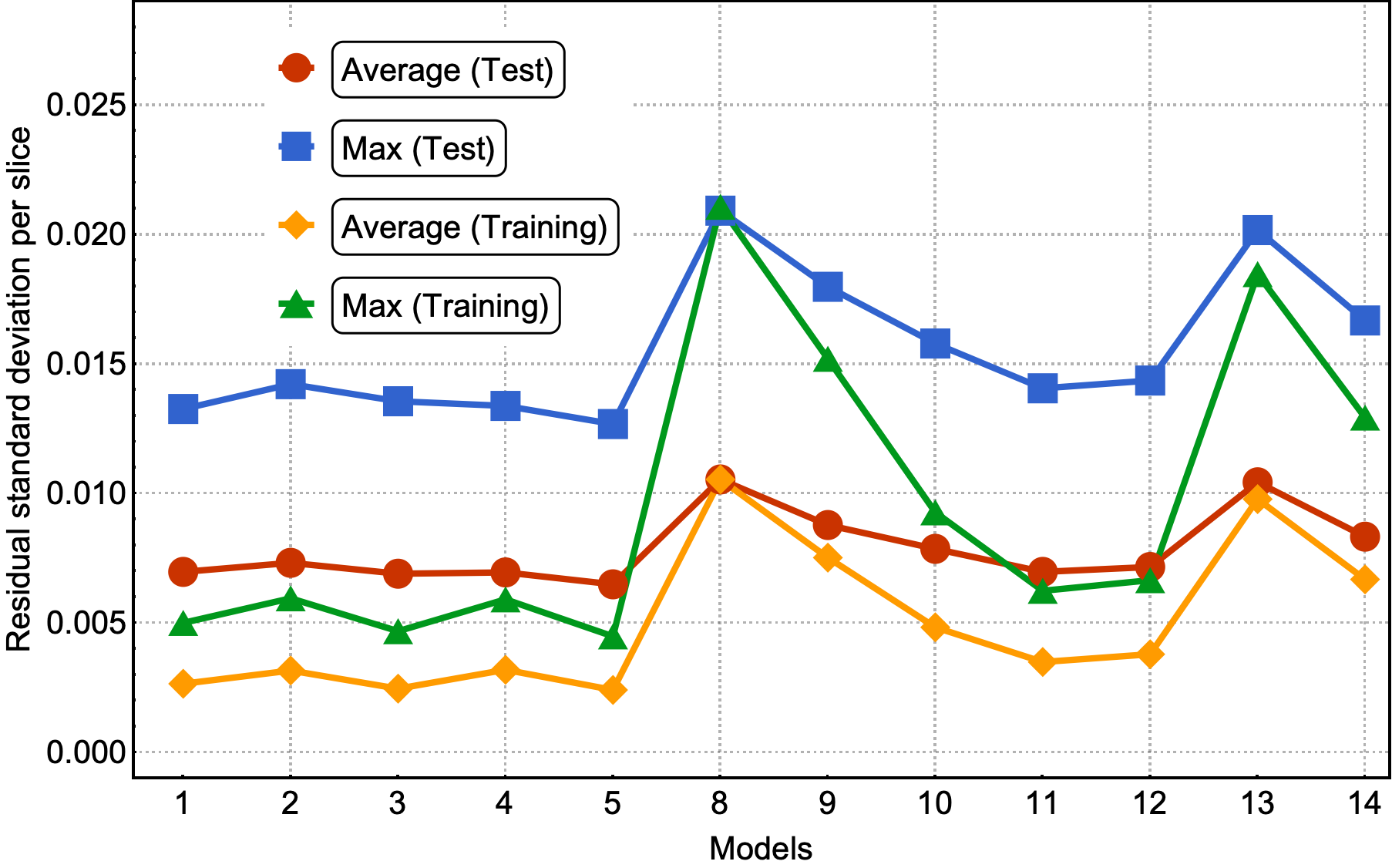}
\label{fig:Plot_s}}
\caption{\textbf{a)} For each model, we show the average and maximum over the residual mean value per slice. \textbf{b)} For each model, we show the average and maximum over the residual standard deviation per slice (see Fig. \ref{fig:Slice-plot}). This was done for the test and training set.} \label{fig:Plot_ms}
\end{figure}

\begin{figure}[hbtp]
\centering
\subfigure[]{
\includegraphics[width=3.1in]{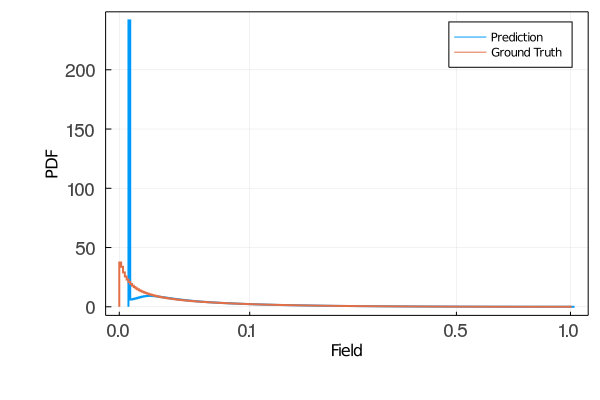}
\label{fig:PDF1}}
\centering
\subfigure[]{
\includegraphics[width=3.1in]{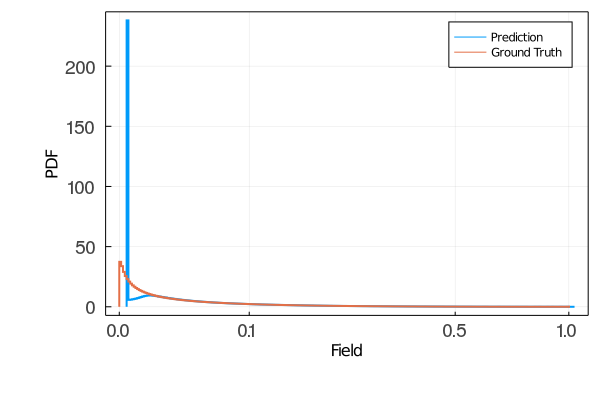}
\label{fig:PDF2}}
\subfigure[]{
\includegraphics[width=3.1in]{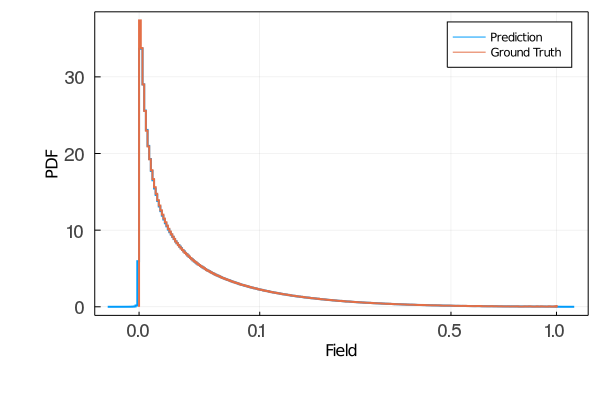}
\label{fig:PDF3}}
\subfigure[]{
\includegraphics[width=3.1in]{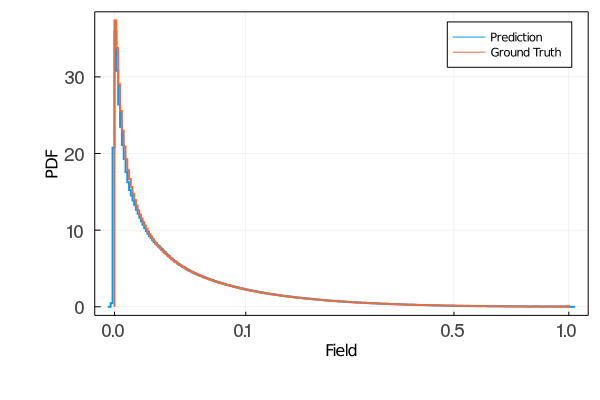}
\label{fig:PDF4}}
\caption{PDF of field obtained via NN (blue) and ground truth (red) in the case of training using MSE, for \textbf{a)} model 2 and \textbf{b)} model 3 and for training using MAE, for \textbf{c)} model 11 and \textbf{d)}  model 12. When using MSE (\textbf{a} and \textbf{b}) the NN predicts zero field values instead of low non-zero field values as the predicted PDF has a larger peak in zero than the ground truth PDF, and a smaller PDF for small non-zero field values compared with the ground truth PDF. When training using MAE (\textbf{c} and \textbf{d}) the prediction and ground truth PDFs overlap completely.} \label{fig:PDF}
\end{figure}

In Fig. \ref{fig:Plot_ms} we  plotted the average and maximum over the residual mean value per slice (see Fig. \ref{fig:Plot_m}) and the residual standard deviation per slice (see Fig. \ref{fig:Plot_s}) for each model's test and training sets. Notice that in this approach, by slicing the residual values and computing the average residual over the set of slices, we are giving equal weight to each mean residual per slice and, therefore, compensating for the imbalance in frequency of low and high value pixels.
An interesting feature from using MSE or MAE comes from the PDF of the field values. Training using MAE makes the PDF prediction quite accurate as the prediction completely overlaps with the ground truth (see Fig. \ref{fig:PDF}). In comparison, when training with MSE, the PDF is not as good and the overlap between ground truth and prediction is not complete.  There is a mismatch for low field values in the sense that the NN does not predict low non-zero field values correctly. Thus we recommend using MAE to avoid this issue.

% \textcolor{red}{jag:You never discussed the histograms of the values, which are critical (figure 8 I think) You need to write a results and discussion section that illustrates what you did and explains it a bit.... The histograms (PDF's) are discussed in line 574 . The true vs prediction plots (Fig 8) are discussed in line 406.}

% \textcolor{red}{jag:You also need to explain the places large absolute and relative errors occur. You can be qualitative and say that you will investigate these more in future work. Has been added to line 403}

%\section{Results and Discussion}
\section{Discussion}

% \textcolor{red}{jps: new separated "Discussion" section. Moved some of the text form the Conclusions section here.}

In large-scale mechanistic simulations of biological tissues, calculations of the diffusion of molecular species can be a significant fraction of the total computational cost. Because biological responses to concentrations  often have a stochastic overlay, high precision may not be essential in these calculations %[????Unfortunately concentrations also vary over 4 over of magnitude range and the relative accuracy may be important, but we won't say this????]
Because NN surrogate estimates are significantly faster than the explicit calculation of the steady-state diffusion field for a given configuration of sources and sinks, an effective NN surrogate could greatly increase the practical size of simulated tissues, e.g., in cardiac simulations \cite{kerckhoffs2007coupling,sundnes2014improved}, cancer simulations \cite{bruno2020progress} and orthopedic simulations \cite{erdemir2007model}. In our case, using a NVIDIA Quadro RTX 6000, each diffusion solution is about 1000 times faster using the trained NN solver compared to the Julia code. 
%This number is only valid for the case...

In order to decide if this acceleration is useful, we have to consider how long it takes to run the direct simulation, how long the NN takes to train and how long it takes to execute the NN once it has been trained \cite{fox2019learning}.
If each diffusion calculation takes $\delta$ seconds to run,  conducting $N$ calculations directly takes $t_{direct} = N\delta$. If each neural network surrogate takes $\epsilon$ seconds to run, and the number of replicas in the training set is $M$ and the training time is $E$, the total time for the neural network simulation is the time to generate the training set, the training time plus the simulation time, $t_{neuro} = M\delta + E + N \epsilon$. To estimate these times, we ran $20000$ explicit simulations in Julia, which took approximately 6 hours and 30 minutes, yielding roughly $1.16 \text{s}$ each. The NN training time was $12$ hours on average. While the speedup for an individual simulation is $\delta/\epsilon \approx 1000$, the ratio $\tau_{neuro}/\tau_{direct}$ must be smaller than 1 in order to have a useful acceleration. Equating this ratio to $1$ and solving for $N$ yields
\begin{equation}
N_{min} = \frac{M+E/\delta}{1-\epsilon/\delta} \approx M + \frac{E}{\delta} \; . \label{eq:minRep}
\end{equation}
$N_{min}$ gives the number of replicas necessary for the total time using the NN to be the same as the direct calculation.
Of course, the exact times will depend on the specific hardware used for the direct and NN calculations. In our case, from Eq. \eqref{eq:minRep} we obtain that $N_{min} \approx 57300$,  we would need to use the neural network more than $57300$ times for the total time using the NN to be faster than the direct calculation. 
%Based on an uncertainty in the ratio $\delta/\epsilon$ of x 10 / 10 the uncertainty in the number of replicas for you to save time is A to B. [Also need to consider whether you are on a GPU or CPU and the number of cores, etc.... to get a cost per core]. 
Thus the NN acceleration is primarily useful in simulations that will be run many, many times for the \textbf{specific situation} for which the NN is appropriate. Consider for example if you wish to include a variable number of sources, different lattice sizes, different dimensionalities (e.g., 3D) and boundary conditions. The more general the NN the more training data it will require, th elonger traingin will take, and the slower the individual NN calculations will be. Currently virtual-tissue simulation studies often run thousands to tens of thousands of replicas and each replica often takes tens of  minutes to tens of hours to run. This computational cost makes detailed parameter identification and uncertainty quantification impractical, since  simulations often have dozens of parameters to explore. If using a NN-based diffusion solver accelerated these simulations by 100$\times$ it would permit practical studies with  hundreds of thousands to millions of replicas, greatly expanding the feasible exploration of parameter space for parameter identification and uncertainty quantification. 
%Say whether this number of replicas makes it attractive or not. Probably it makes it only interesting if a single NN can do lots of situations (e.g. an arbitrary number of sources and sinks). However, the more situations we want a single NN to handle,  Here the number of replicas in the training set becomes critical and we will explore this in future work. 

\section{Conclusions}

Neural networks provide many possible approaches to generating surrogate diffusion solvers. Given the type of problem setting, we were interested in a neural network that could predict the stationary field.
%and detect the sources. 
We considered a deep convolutional neural network, an autoencoder and their combination. We considered two loss functions, \textit{viz.} mean squared error and mean absolute error. We considered different hyperparameters for dropout and an exponential weight to compensate the under-sampling of high field values. The exponential weight also helped reduce overfitting as shown in Fig. \ref{fig:true-output}.

The range of scientific and engineering applications for diffusion solvers is very broad. Depending on the specific application, the predictions by the neural network will have to meet a specific set of criteria quantified in the form of statistical estimators
%may have different criteria for what constitutes an acceptable neural network approximation 
(\textit{e.g}. mean error, max error, percentiles, mean relative error, \textit{etc}.). In this paper we studied several reasonable error metrics, namely, mean residual, maximum residual, 99-Percentile residual, mean relative residual, mean weighted residual and the weighted standard deviation residual. The last two metrics compensate for the low frequency of  high field values, ones that usually occur in small regions around sources.
%[list and explain why the choices we made are reasonable]. 
%We evaluated two network architectures and their superposition [discuss] and investigated how the choice of hyperparameters affected the different error metrics we considered individually and collectively. 
%We found that the the convolutional neural network was able to detect the sources, while the autoencoder was able to predict the field and also detect the sources although to a less extent. 
The autoencoders are commonly used in generative models which is applicable, as we have shown here, to the case of a diffusion surrogate. The field predictions are accurate on all the metrics we considered. This is appears to be due to collapsing the input into a one-dimensional vector and then decoding back to the initial size, which forces the network to learn the relevant features \cite{kingma2019introduction}. 
%The deep convolutional neural network generated bad metrics results since, as mentioned, it only detects the sources. 
%[You need to say that you evaluated different error functions and what you found and why] [when does the surrogate tend to work well(how defined?] and when doesn't it?] 
While some models had high errors across all metrics, no single model had the smallest error for all error metrics. Different networks and hyperparameters were optimal for different metrics, \textit{e.g.} model 5 had the lowest mean residual, whereas model 9 yielded relatively good results on all metrics. Model 9 uses both neural networks with the dropout values for the deep convolutional network were set to $D_{1,2}=0.4$, and for the autoencoder to $D_{3,4}=0.1$. 
%These values are on the range of what is recommended by the machine learning community.
The weight hyperparameter was set to $100$. Recall that large weight hyperparameter values make the loss function weight high field values over low field values. This is important since the largest absolute error happens close to sources and close to boundaries because of the under-representation of these kinds of configurations. We also noticed that this choice reduced the overfitting as was shown in Fig. \ref{fig:true-output}.

Additionally, we tested several loss function. Here we reported the results using mean squared error and mean absolute error. We noticed two key differences. With MSE the weighted standard deviation (see Fig. \ref{fig:Slice-plot}) is smaller than for MAE for the training set. However, for the test set, the results for both loss functions are comparable. This difference between training and test sets suggests that MSE is more prone to overfitting the data than MAE. The other key difference is that for the MAE, the predicted field probability function consistently overlapped the ground truth completely, whereas for MSE there is a mismatch in that the NN does not predict low non-zero field values correctly (see Fig. \ref{fig:PDF}). Therefore, we recommend using MAE as the loss function for surrogate calculations where the field values are well bounded, as we have shown it produces better predictions than MSE. The autoencoder (NN 2) is capable of approximating the diffusion field on its own, the convolutional network (NN 1) is not. However, if we use the two networks together we find that the prediction is more accurate than NN 2 alone. 
%The autoencoder NN is crucial for this problem, while the superposition with the deep convolutional network improves the results.
%Furthermore, to have good results over the different metrics considered, low dropout values should be used for the autoencoder. 
%The weight hyperparameter helps both in compensating the under-sampled high field values and reducing overfitting. 
%Finally, while MAE contributes to the latter, it also improves the predicted field values probability density function.

% We investigated:
% 1) two network architectures
% 2) several error functions
% 3) Several sets of hyperparameters
% 4) Several different metrics for agreement between the estimate and ground truth

% We found.....

% Were there specific geometric situations that caused problems....? Sources near each other, near edge, big gradients,...

% this suggests that...

% Based on this the next thing we will do is....

These encouraging results suggest that we should pursue NN surrogates for acceleration of simulations in which the solution to the diffusion equation contributes a considerable fraction of the total computational cost. An effective NN diffusion solver surrogate would need to be able to solve diffusion fields for arbitrary sources and sinks in two or three dimensions with variable diffusivity, a much higher dimensional set of conditions than the two circular sources in a uniform two-dimensional square domain that we investigated in this paper. A key question will be the degree to which NNs are able to generalize, \textit{e.g.} from $n$ sources to $n+1$ sources or from circular sources to more complex shapes. In addition, here we only considered absorbing boundary conditions, ultimately mixed boundary conditions are desirable. It is unclear if the best approach would be a single NN capable of doing multiple boundary conditions, or better to develop unique NNs for each boundary condition scenario. We will consider these extensions in future work.

%\textbf{[]What else would you try next? How could you define what situations challenge the accuracy of the simulations]}
%\textbf{I would also mention that solving Navier stokes or mechanical deformation in continuum elasticity requires the solution of diffusion eqn. This this work is an important step towards solving these more complex PDEs}

\section{Acknowledgements}
This research was supported in part by Lilly Endowment, Inc., through its support for the Indiana University Pervasive Technology Institute.

This work is partially supported by the National Science Foundation (NSF) through awards nanoBIO 1720625, CINES 1835598 and Global Pervasive Computational Epidemiology 1918626 and Cisco University Research Program Fund grant 2020-220491.

This work is partially supported by the Biocomplexity Institute at Indiana University, National Institutes of Health, grant NIGMS R01 GM122424

%     %note the cite above. That is an in text citation. By putting citations in your bibliography, you can cite in text simply using the \cite{} command and putting the number of the citation in the {} 

\newpage

\bibliographystyle{unsrt}  
\bibliography{refs}

\newpage
\appendix

% jps: reset figure numbers and change the label
%\renewcommand{\thefigure}{Supplement Figure \arabic{figure}}
% \setcounter{figure}{0}
\newcommand{\beginsupplement}{%
        \setcounter{table}{0}
        \renewcommand{\thetable}{S\arabic{table}}%
        \setcounter{figure}{0}
        \renewcommand{\thefigure}{S\arabic{figure}}%
     }
\beginsupplement
\section{Supplementary material: Probability density function per slice} \label{app:PDF_Slices}
% \textcolor{red}{Should restart the figure numbers as 1 and name the figures "Supplement Figure \#.}

In this section we show the PDF's per slice as described in the main text. We took $20$ slices of size $0.05$ in the direction $y=x$ for the plots shown in Fig. \ref{fig:true-output}. We then compute the mean residual and standard deviation per slice. For each slice, we have also plotted a Gaussian distribution (red curve) for guidance purposes which has mean and standard deviation set to the mean residual and standard deviation per slice, respectively.

\begin{figure}[hbtp]
\centering
% \subfigure[Model 11 Test]{
\includegraphics[width=8.4in, angle = -90]{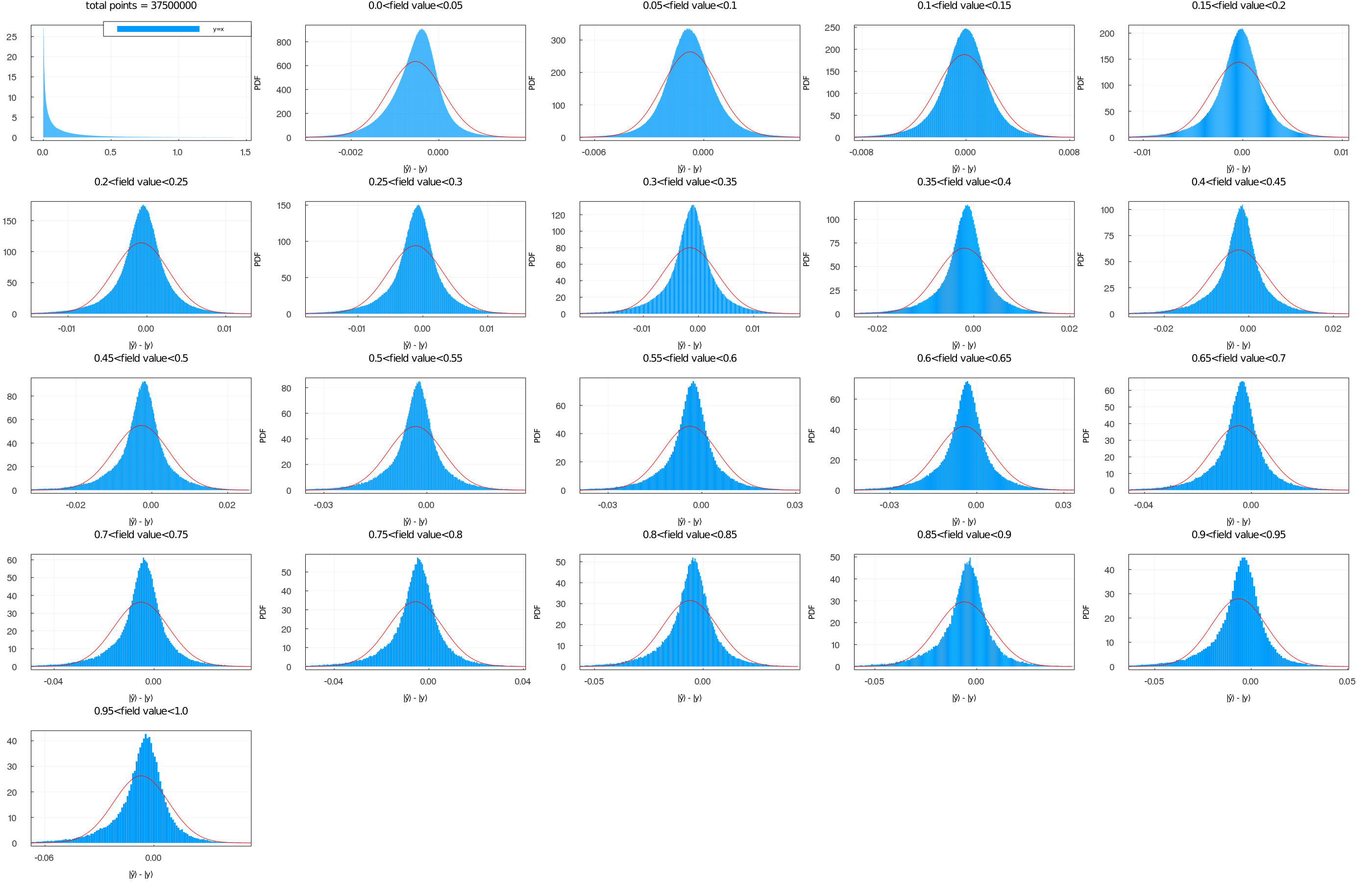}
% \label{fig:11-test}}
\caption{Model 1. PDF over all field values and PDF over slices. The red curve corresponds to a Gaussian distribution centered at the PDF mean value and with standard deviation equal to that of the PDF.} \label{fig:Hist_Plot_1}
\end{figure}

\begin{figure}[hbtp]
\centering
% \subfigure[Model 11 Test]{
\includegraphics[width=8.4in, angle = -90]{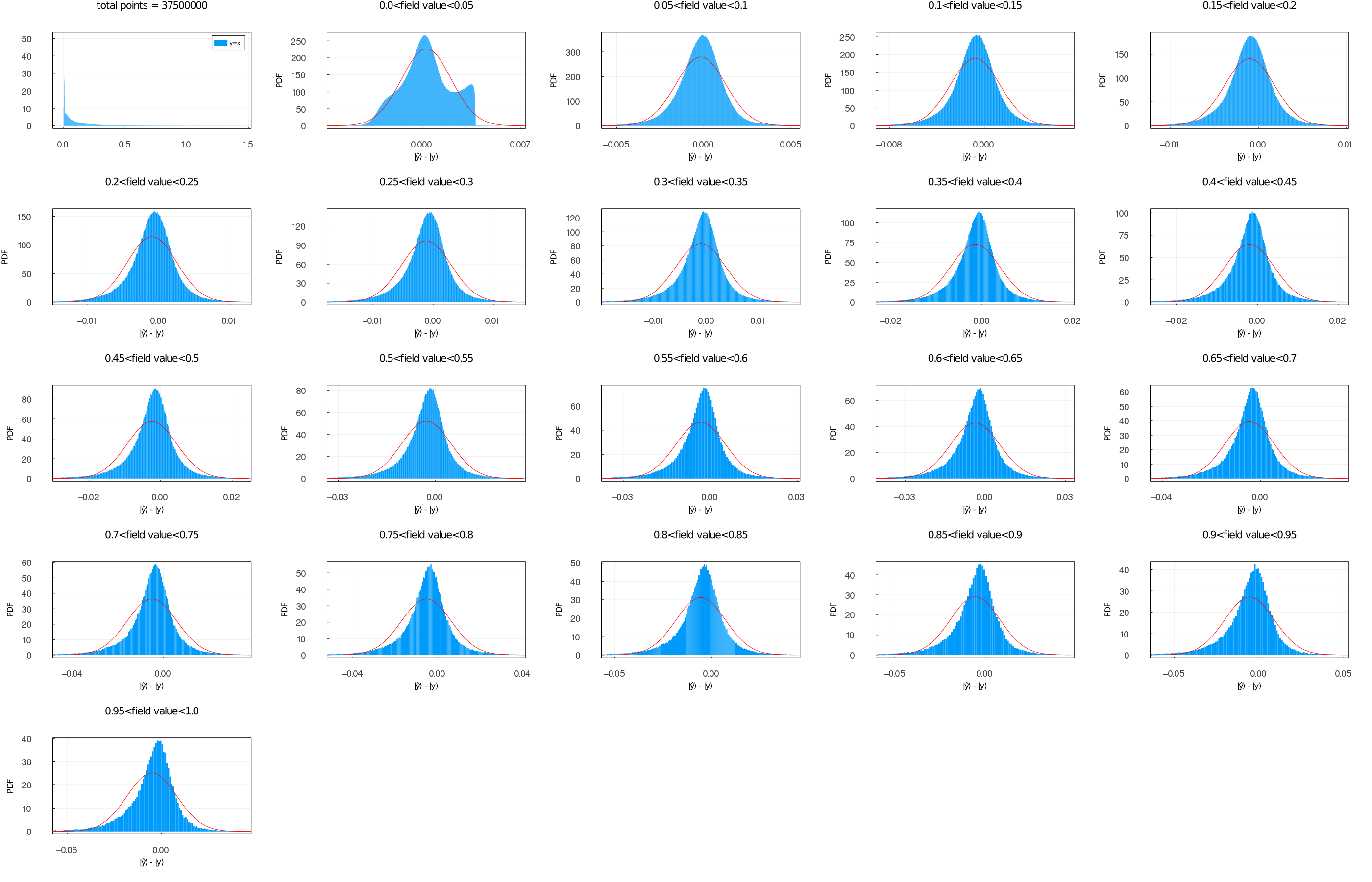}
% \label{fig:11-test}}
\caption{Model 2. PDF over all field values and PDF over slices. The red curve corresponds to a Gaussian distribution centered at the PDF mean value and with standard deviation equal to that of the PDF.} \label{fig:Hist_Plot_2}
\end{figure}

\begin{figure}[hbtp]
\centering
% \subfigure[Model 11 Test]{
\includegraphics[width=8.4in, angle = -90]{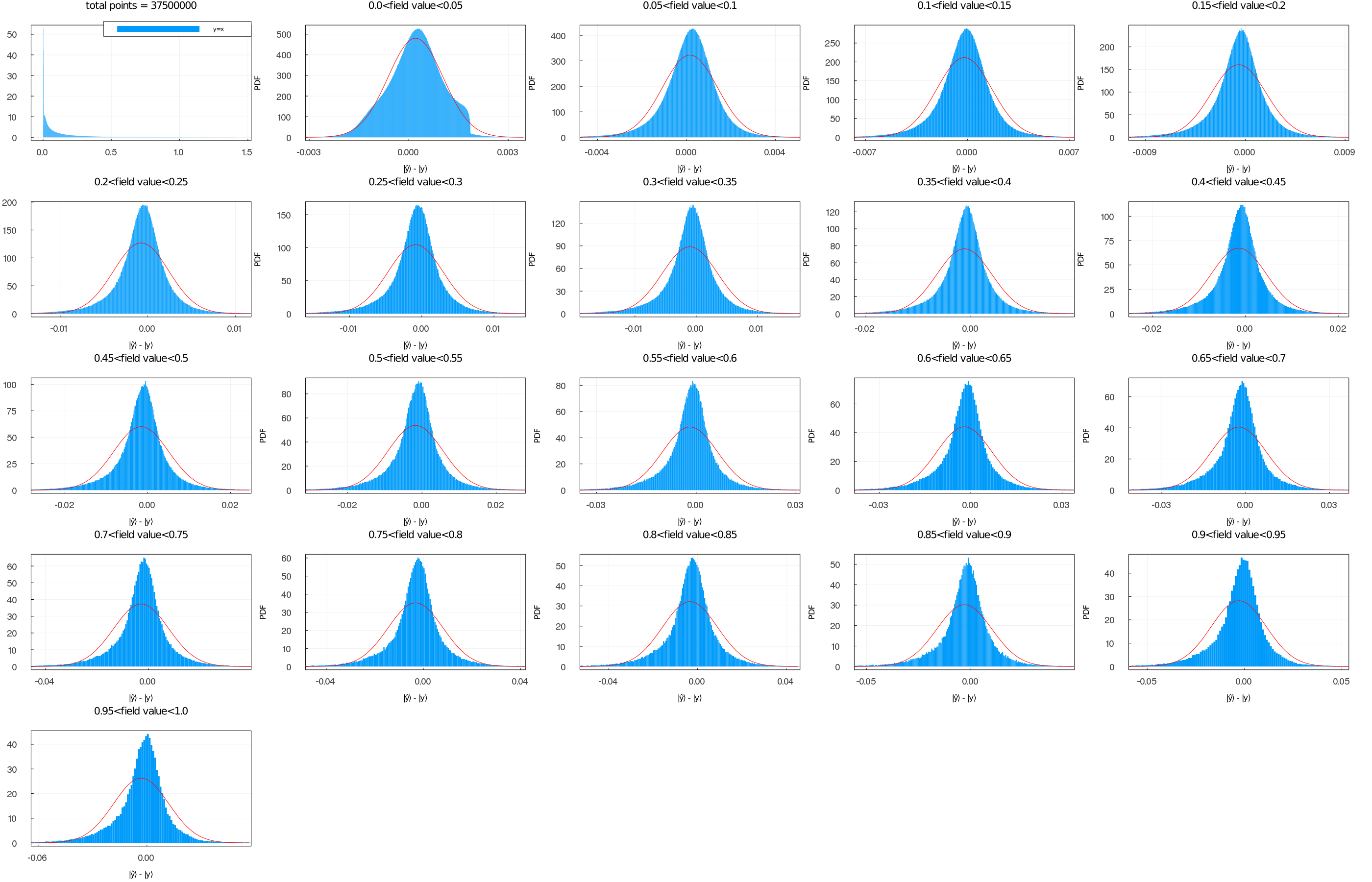}
% \label{fig:11-test}}
\caption{Model 3. PDF over all field values and PDF over slices. The red curve corresponds to a Gaussian distribution centered at the PDF mean value and with standard deviation equal to that of the PDF.} \label{fig:Hist_Plot_3}
\end{figure}

\begin{figure}[hbtp]
\centering
% \subfigure[Model 11 Test]{
\includegraphics[width=8.4in, angle = -90]{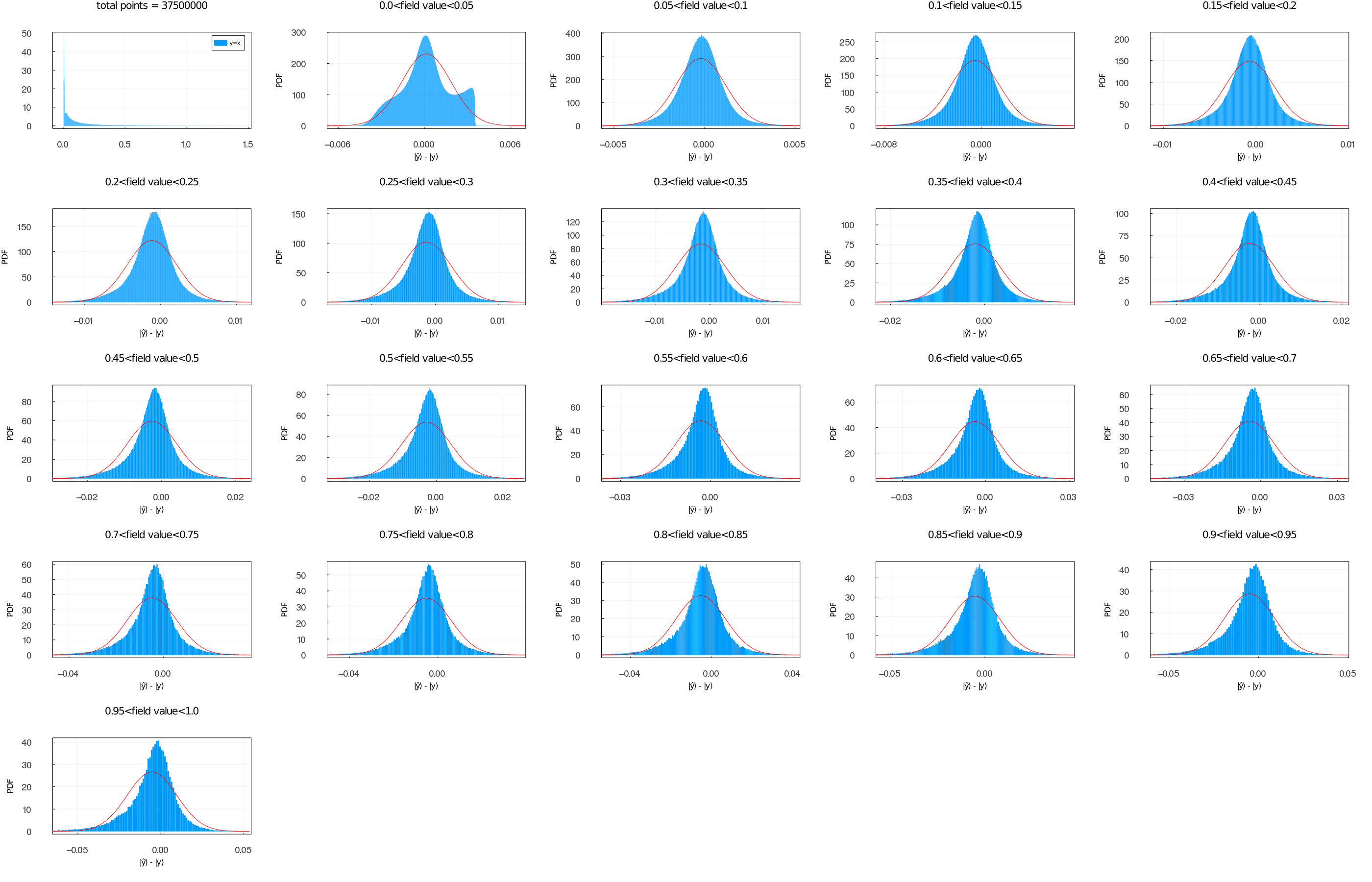}
% \label{fig:11-test}}
\caption{Model 5. PDF over all field values and PDF over slices. The red curve corresponds to a Gaussian distribution centered at the PDF mean value and with standard deviation equal to that of the PDF.} \label{fig:Hist_Plot_5}
\end{figure}

\begin{figure}[hbtp]
\centering
% \subfigure[Model 11 Test]{
\includegraphics[width=8.4in, angle = -90]{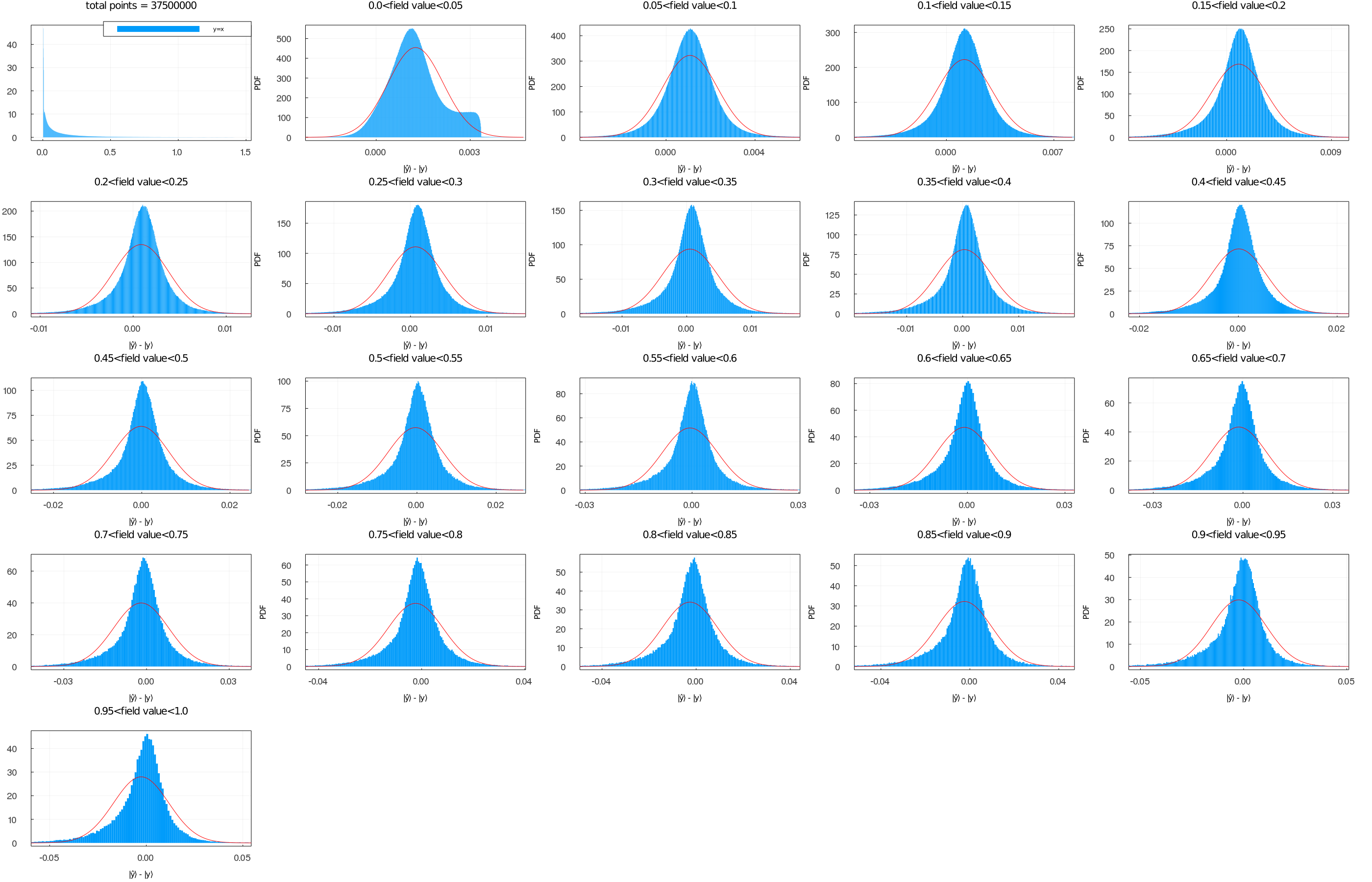}
% \label{fig:11-test}}
\caption{Model 6. PDF over all field values and PDF over slices. The red curve corresponds to a Gaussian distribution centered at the PDF mean value and with standard deviation equal to that of the PDF.} \label{fig:Hist_Plot_6}
\end{figure}

\begin{figure}[hbtp]
\centering
% \subfigure[Model 11 Test]{
\includegraphics[width=8.4in, angle = -90]{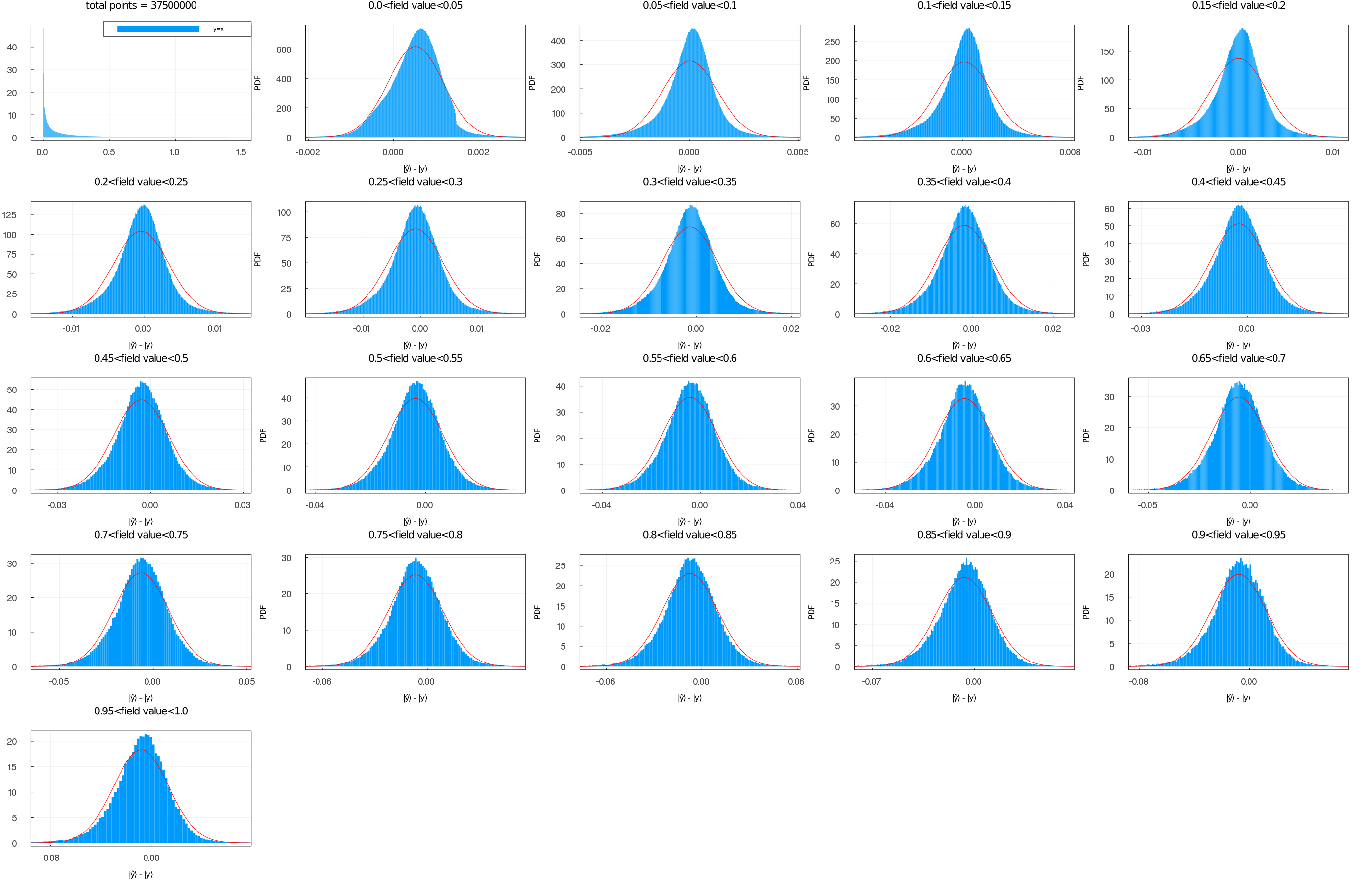}
% \label{fig:11-test}}
\caption{Model 8. PDF over all field values and PDF over slices. The red curve corresponds to a Gaussian distribution centered at the PDF mean value and with standard deviation equal to that of the PDF.} \label{fig:Hist_Plot_8}
\end{figure}

\begin{figure}[hbtp]
\centering
% \subfigure[Model 11 Test]{
\includegraphics[width=8.4in, angle = -90]{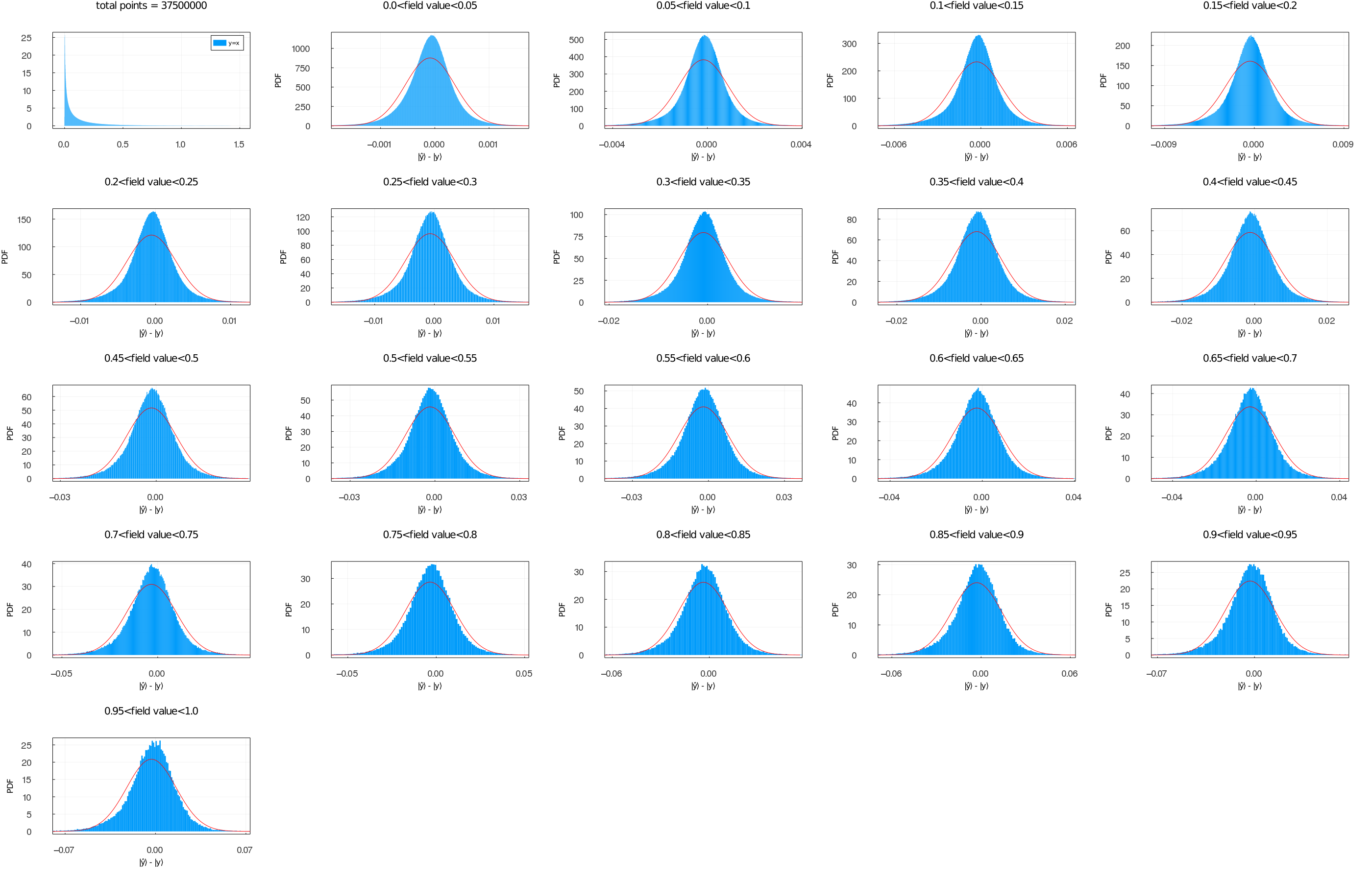}
% \label{fig:11-test}}
\caption{Model 9. PDF over all field values and PDF over slices. The red curve corresponds to a Gaussian distribution centered at the PDF mean value and with standard deviation equal to that of the PDF.} \label{fig:Hist_Plot_9}
\end{figure}

\begin{figure}[hbtp]
\centering
% \subfigure[Model 11 Test]{
\includegraphics[width=8.4in, angle = -90]{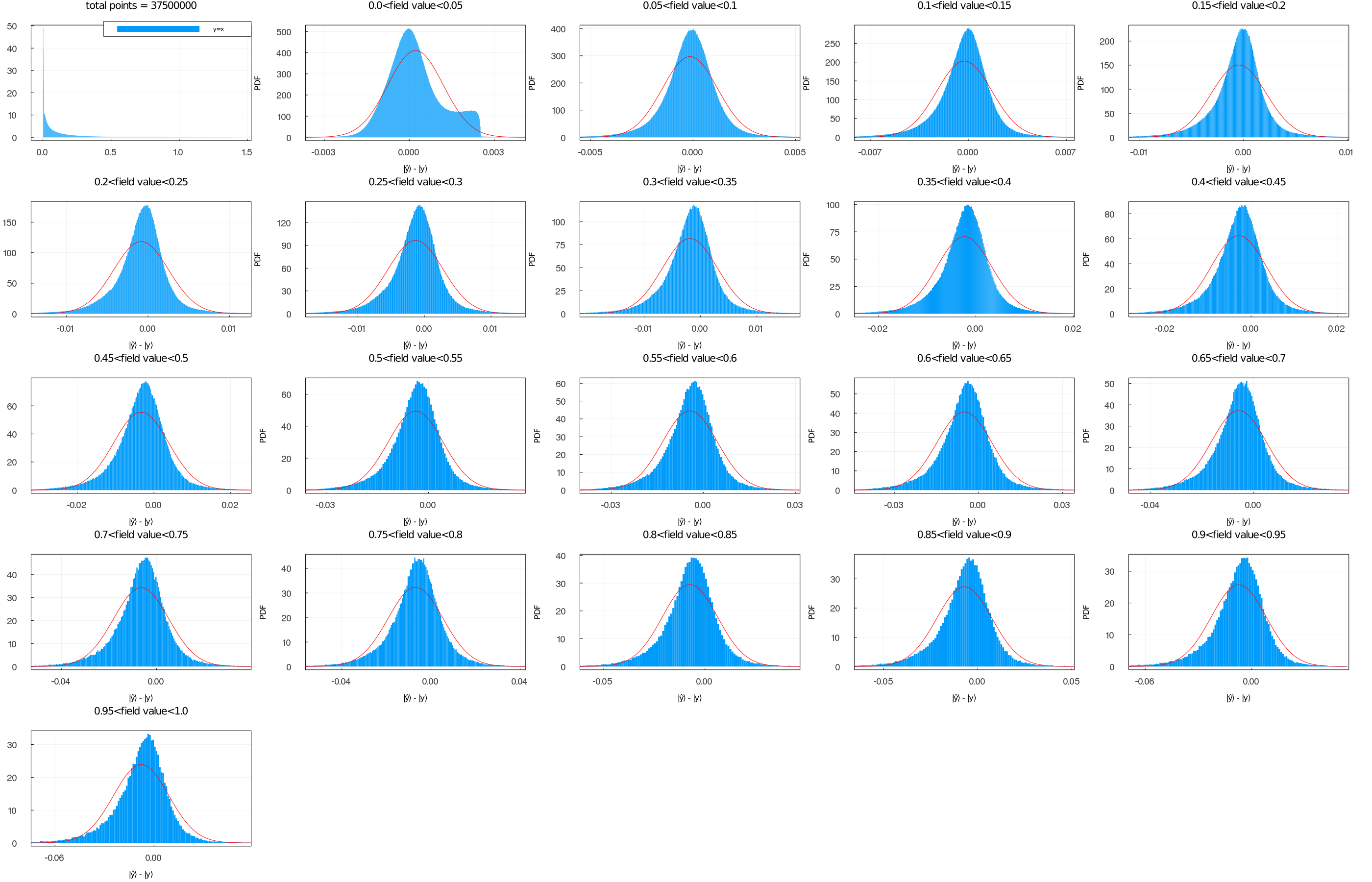}
% \label{fig:11-test}}
\caption{Model 10. PDF over all field values and PDF over slices. The red curve corresponds to a Gaussian distribution centered at the PDF mean value and with standard deviation equal to that of the PDF.} \label{fig:Hist_Plot_10}
\end{figure}

\begin{figure}[hbtp]
\centering
% \subfigure[Model 11 Test]{
\includegraphics[width=8.4in, angle =-90]{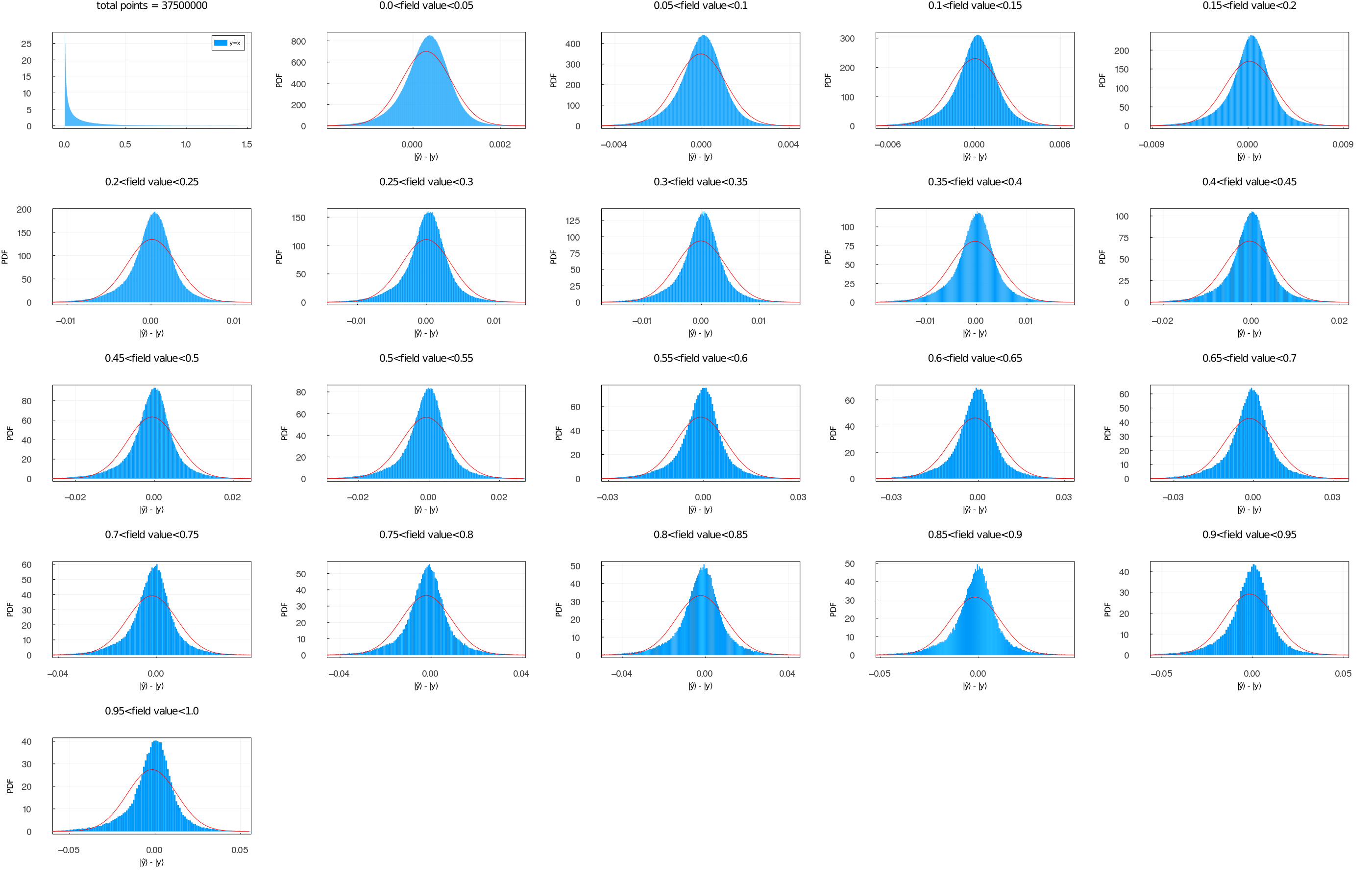}
% \label{fig:11-test}}
\caption{Model 11. PDF over all field values and PDF over slices. The red curve corresponds to a Gaussian distribution centered at the PDF mean value and with standard deviation equal to that of the PDF.} \label{fig:Hist_Plot_11}
\end{figure}

\begin{figure}[hbtp]
\centering
% \subfigure[Model 11 Test]{
\includegraphics[width=8.4in, angle =-90]{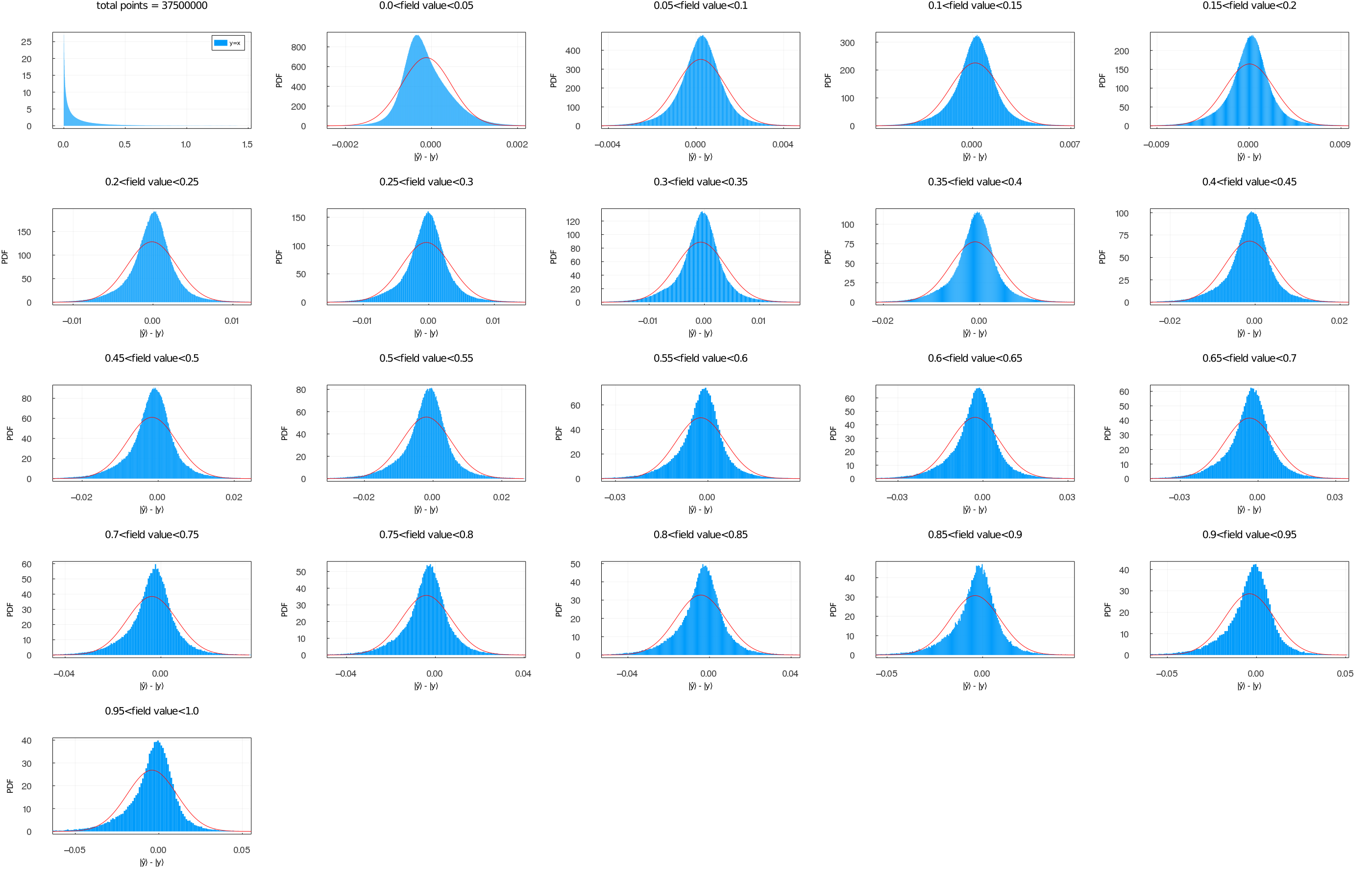}
% \label{fig:11-test}}
\caption{Model 12. PDF over all field values and PDF over slices. The red curve corresponds to a Gaussian distribution centered at the PDF mean value and with standard deviation equal to that of the PDF.} \label{fig:Hist_Plot_12}
\end{figure}

\begin{figure}[hbtp]
\centering
% \subfigure[Model 11 Test]{
\includegraphics[width=8.4in, angle =-90]{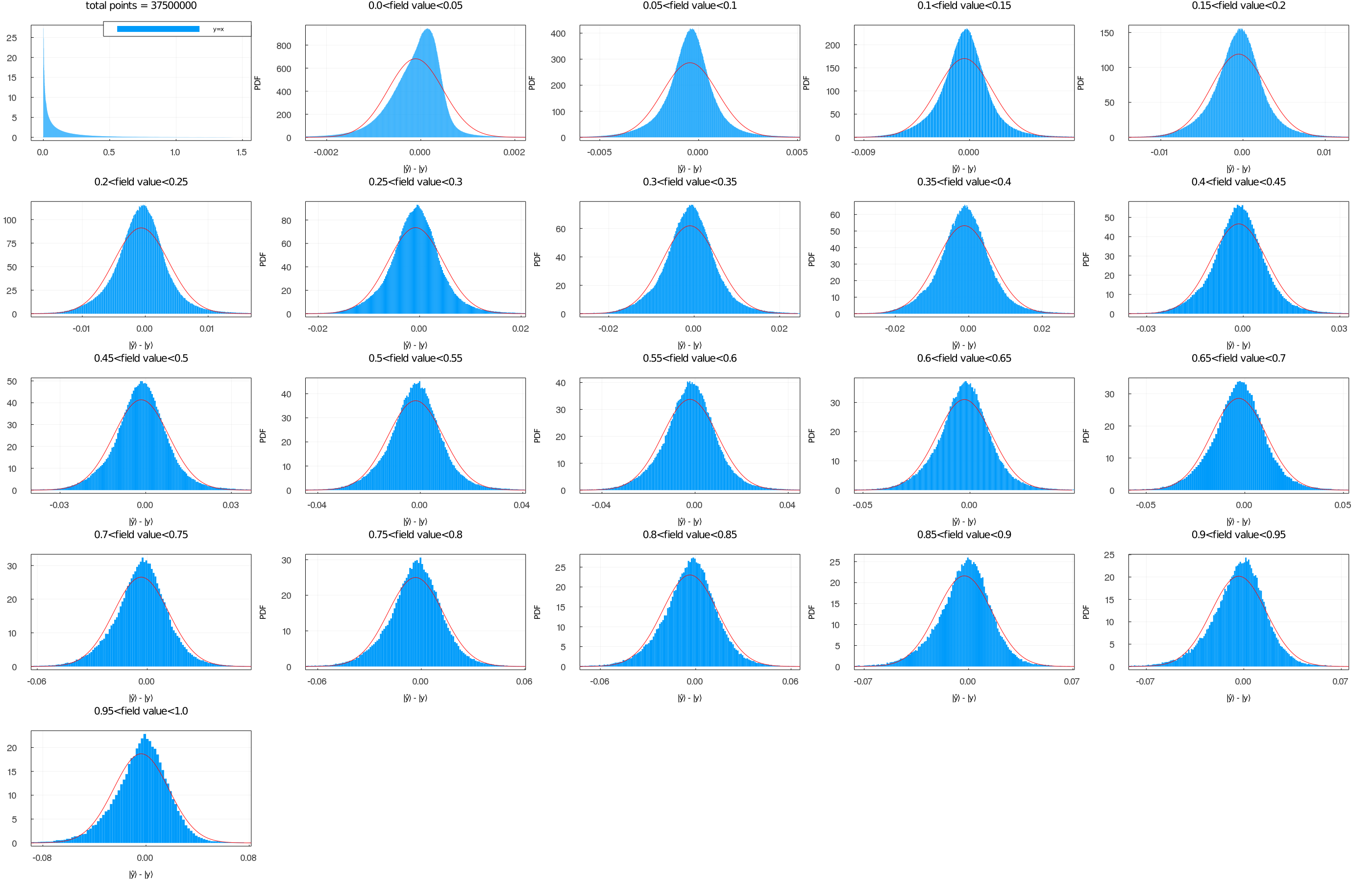}
% \label{fig:11-test}}
\caption{Model 13. PDF over all field values and PDF over slices. The red curve corresponds to a Gaussian distribution centered at the PDF mean value and with standard deviation equal to that of the PDF.} \label{fig:Hist_Plot_13}
\end{figure}

\begin{figure}[hbtp]
\centering
% \subfigure[Model 11 Test]{
\includegraphics[width=8.4in, angle =-90]{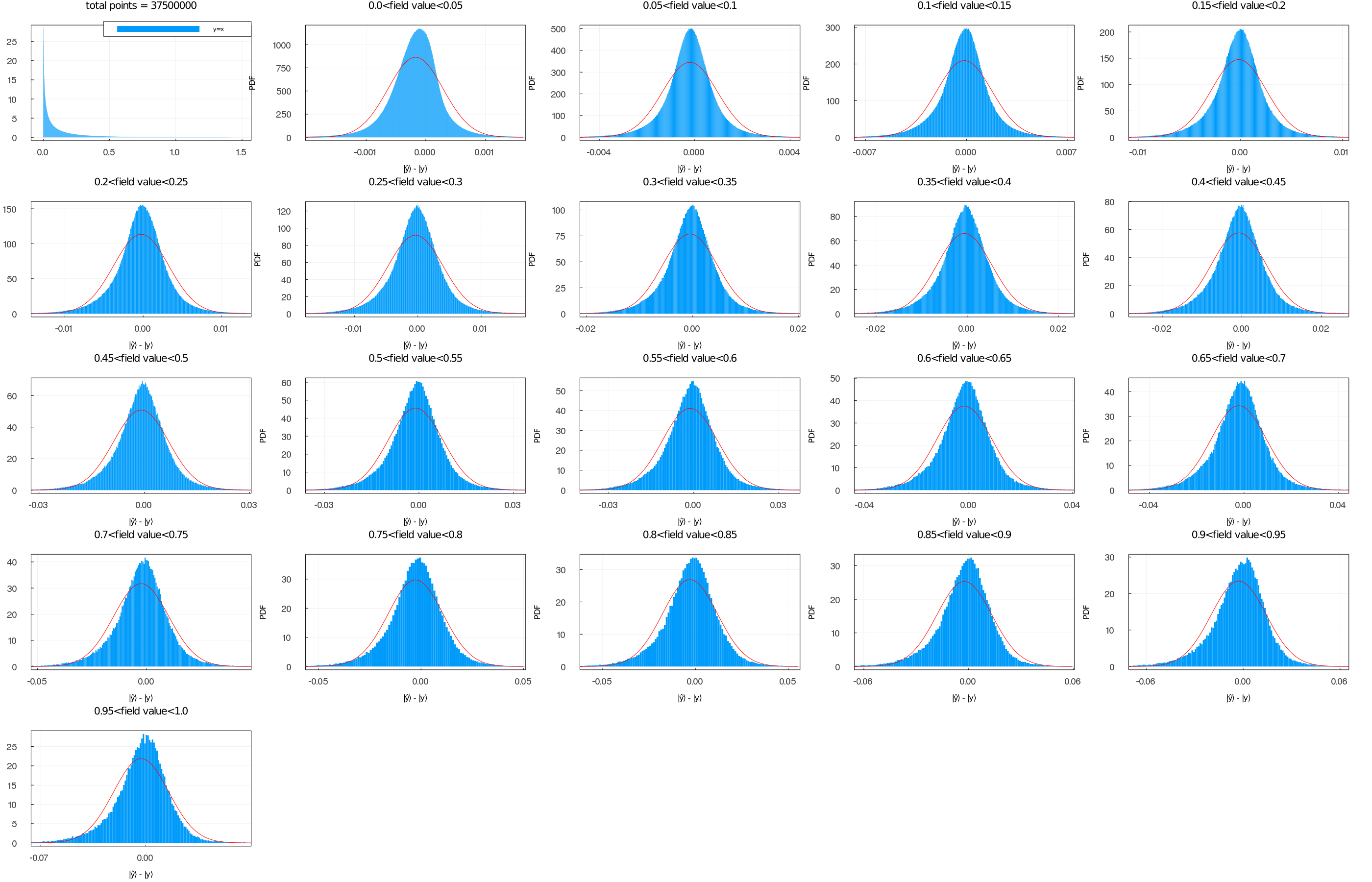}
% \label{fig:11-test}}
\caption{Model 14. PDF over all field values and PDF over slices. The red curve corresponds to a Gaussian distribution centered at the PDF mean value and with standard deviation equal to that of the PDF.} \label{fig:Hist_Plot_14}
\end{figure}

\end{document}